\documentclass[a4paper,UKenglish,cleveref, autoref, thm-restate]{lipics-v2021}
\usepackage{amsthm}
\usepackage{array}
\usepackage{alltt}
\usepackage{mathtools}
\usepackage{amssymb}
\usepackage{xfrac}
\usepackage{cleveref}
\usepackage{tikz}
\usepackage[linesnumbered]{algorithm2e}
\usepackage{stmaryrd}
\usepackage{algorithmicx}
\usepackage{adjustbox}

\hideLIPIcs 
\usetikzlibrary{arrows,automata}
\usetikzlibrary{automata,positioning,decorations.pathreplacing}

\newtheorem{fact}[theorem]{Fact}

\newcommand{\B}{\mathbb B}
\newcommand{\Nat}{\mathbb N}
\newcommand{\partit}{\pi}
\renewcommand{\epsilon}{\varepsilon}

\newcommand{\pijl}[1]{\mathrel{\text{$\xrightarrow{\smash{\text{\raisebox{-1pt}{$#1$}}}}$}}}

\newcommand{\sem}[1]{\llbracket #1 \rrbracket}
\newcommand{\equivclass}[1]{%
  S/{#1}%
}
\newcommand{\classes}[1] {\equivclass{#1}}
\newcommand{\bisim}{\leftrightarroweq}

\newcommand{\spl}{\textit{split}}

\newcommand\simby{%
	  \mathrel{\ooalign{\raise0.15em\hbox{\scalebox{0.7}{$\rightarrow$}}\cr%
	  \lower0.1em\hbox{\scalebox{0.8}{$\--$}}}}}
\newcommand\simueq{%
	  \mathrel{\ooalign{\raise0.15em\hbox{\scalebox{0.7}{$\rightarrow$}}\cr%
	  \lower0.1em\hbox{\scalebox{0.7}{$\leftarrow$}}}}}

\newcommand\simbyaux{%
	  \mathrel{\ooalign{\raise0.15em\hbox{\scalebox{0.7}{$\rightsquigarrow$}}\cr%
	  \lower0.1em\hbox{\scalebox{0.8}{$\--$}}}}}
	  
\renewcommand{\leq}{\leqslant}
\newcommand{\sba}[2]{\simbyaux_{#1}^{#2}}
\newcommand{\splpairs}{\delta}

\newcommand{\dirdistsingle}{\overrightarrow{\Delta}}
\newcommand{\dist}{\Delta}

\newcommand{\dia}[1]{\langle {#1} \rangle}
\newcommand{\ttrue}{\textit{tt}}
\newcommand{\ffalse}{\textit{ff}}

\newcommand{\bigO}{\mathcal{O}}
\newcommand{\F}{\mathcal{F}}
\newcommand{\X}{\mathcal{X}}
\newcommand{\negF}{\mathcal{F}}

\newcommand{\depth}[1]{\mathit{d}_{\diamond}(\mkern1mu {#1}\mkern1mu)}
\newcommand{\size}[1]{|#1|}   
\newcommand{\negdepth}[1]{\mathit{d}_{\neg}(\mkern1mu {#1}\mkern1mu)}
\newcommand{\depthf}[1]{\mathit{d}_{\diamond}}
\newcommand{\sizef}[1]{|\cdot|}
\newcommand{\negdepthf}[1]{\mathit{d}_{\neg}}

\newcommand{\prop}{\mathit{Prop}}
\renewcommand{\flag}{\mathtt{false}}
\newcommand{\sat}{\mathtt{sat}}
\newcommand{\unsat}{\mathtt{unsat}}
\newcommand{\bottom}{\bot}
\newcommand{\C}{\mathcal{C}}
\newcommand{\init}{\mathtt{init}}
\newcommand{\traces}{\textit{Tr}}
\newcommand{\truths}{\textit{Truths}}
\newcommand{\concat}{\cdot}

\newcommand{\calA}{\mathcal{A}}
\newcommand{\calB}{\mathcal{B}}

\title{Computing minimal distinguishing Hennessy-Milner formulas is NP-hard, 
but variants are tractable}

\author{Jan Martens}{Eindhoven University of Technology,
The Netherlands}{j.j.m.martens@tue.nl}{https://orcid.org/0000-0003-4797-7735}{AVVA project NWO 612.001.751/TOP1.17.002}
\author{Jan Friso Groote}{Eindhoven University of Technology,
The Netherlands}{j.f.groote@tue.nl}{https://orcid.org/0000-0003-2196-6587}{}

\authorrunning{J. Martens and J.F. Groote} 

\Copyright{Jan Martens and Jan Friso Groote} 

\titlerunning{On computing minimal distinguishing Hennessy-Milner formulas}


\supplementdetails[]{Prototype}{https://github.com/jjmartens/distinguishing-hml} 

\nolinenumbers

\begin{document}
	\maketitle  
   \begin{abstract}
    \noindent%
We study the problem of computing minimal distinguishing formulas
for non-bisimilar states in finite LTSs. We show that this is NP-hard if the
size of the formula must be minimal. Similarly, the existence of a short
distinguishing trace is NP-complete. 
However, we can provide polynomial algorithms, if minimality is formulated
as the minimal number of nested modalities, and it can even be extended by
recursively requiring a minimal number of nested negations. A prototype 
implementation shows that the generated formulas are much smaller than those
generated by the method introduced by Cleaveland. 
  \end{abstract}

\section{Introduction}
Hennessy-Milner Logic (HML)~\cite{hennessy1980observing} can be used to
explain behavioural inequivalence. If two states are not
bisimilar there is a \textit{distinguishing formula} that 
is valid in one state but not in the other. As the reason for 
the states not being bisimilar can be very subtle, such a distinguishing
formula is of great help to pinpoint the cause of the inequivalence. 

Cleaveland~\cite{cleaveland1990} introduces an efficient algorithm to calculate
distinguishing formulas by back-tracking the partition refinement sequence that
decides bisimilarity. He states that the formulas are minimal ``in a precisely
defined sense''. This method is used in the mCRL2 toolset~\cite{mcrl2}. However,
the generated formulas are unexpectedly large. This leads to the question in
which sense distinguishing formulas are minimal and how difficult it is to
obtain them. Similar questions were posed throughout the literature. Some also
questioned the size of the formulas -- in the setting of CTL \cite{BrowneCG88},
others explicitly stated that they were not minimal
\cite{wissmann2022quasilinear,nestmann2022deciding}, and there are even
suggestions that minimisation could be NP-hard \cite{CAAL20}. 


In this work we answer the question by proving that in general calculating
minimal distinguishing Hennessy-Milner formulas is NP-hard. Minimality can
be taken rather broadly, as having a minimal number of symbols, modalities,
or logical connectives. As observed in~\cite{figueira2010size} a distinguishing
formula can be exponential in size. However, as was already noted in 
\cite{cleaveland1990}, when using sharing in the representation of formulas, 
for instance by formulating the distinguishing formula as a set of equations
or a directed acyclic graph,
the representation is polynomial. Calculating a minimal shared
distinguishing formula is NP-complete.

The proof of this result uses a reduction directly from CNF-SAT and the construction
is similar to the construction used by Hunt~\cite{hunt74} where it is shown that
deciding equivalence of acyclic non-deterministic automata is NP-complete.
We show via the NP-hardness of deciding whether there is a distinguishing trace
for an acyclic non-deterministic LTS, that computing minimal HML formulas is also NP-hard. 

As distinguishing formulas are very useful, we are wondering whether a variant
of minimality of distinguishing formulas exists that leads to concise formulas
and that can effectively be calculated. We answer this positively by providing
efficient algorithms to construct distinguishing formulas that are minimal with
respect to the \emph{observation-depth}, i.e., the number of nested modalities.
Within this we can even guarantee in polynomial time that the
\emph{negation-depth}, i.e., the number of nested negations, or equivalently the
number of nested alternations of box and diamond modalities, is minimal. These
algorithms strictly improve upon the method by Cleaveland~\cite{cleaveland1990}.
A prototype implementation of our algorithm shows that our formulas are indeed
much smaller and more pleasant to use. In order to obtain these results we
employ the notions of $k$-bisimilarity~\cite{milner1980calculus} and $m$-nested
similarity~\cite{groote1992structured}.

Distinguishing formulas have been the topic of studies in many papers, more
than we can mention.
A recent impressive work introduces a method to find minimal
distinguishing formulas for various classes of behavioural equivalences~\cite{nestmann2022deciding}. 
The algorithm translates the problem to 
determining the winning region in a reachability game. These games can grow super-exponentially in size.
In the context of distinguishing deterministic finite automata,
an algorithm is given that from a splitting tree
finds pairwise minimal distinguishing words~\cite{smetsers2016minimal}. In a more generalized
setting~\cite{wissmann2022quasilinear, konig2020explaining} a co-algebraic method
is given to generate distinguishing modal formulas.
The notion of distinguishing formulas is also used in the setting with
abstractions for branching bisimilarity~\cite{korver1991computing, geuvers2022apartness}.

This document is structured as follows. In Section~\ref{sec:prelims} the
required preliminaries on LTSs and HML formulas are given. In
Section~\ref{sec:np-hard}, we show that decision problems related to finding
minimal distinguishing formulas are NP-hard. Next, in Section~\ref{sec:dist} we
give a procedure that generates a minimal observation- and negation-depth
formula. Additionally, in this section, we give a partition refinement algorithm
inspired by~\cite{smetsers2016minimal,moore1956gedanken} which can be used to
determine minimal observation-depth distinguishing formulas. In the full
version, an appendix is included containing proofs omitted here due to space
constraints. 

\section{Preliminaries}\label{sec:prelims}
\noindent For the numbers $i,j\in \Nat$, we define $[i,j] = \{ c \in \Nat \mid i
\leq c \leq j\}$, the closed interval from $i$ to $j$.\looseness=-1

\subsection{LTSs, $k$-bisimilarity \& $m$-nested similarity}
\noindent We use Labelled Transition Systems (LTSs) as our behavioural models.
Strong bisimilarity is a widely used behavioural
equivalence~\cite{milner1980calculus, park1981concurrency}, which we define in
the classical inductive way.

\begin{definition}
	A labelled transition system~(LTS) $L=(S, Act, \pijl{})$ is a three-tuple
	containing:
	\begin{itemize}
		\item a finite set of states $S$,
		\item a finite set of action labels $Act$, and
		\item a transition relation ${\pijl{}}\subseteq S\times Act \times S$.
	\end{itemize}
\end{definition}

\noindent We write $s\pijl{a}s'$ iff $(s, a, s')\in
\pijl{}$. We call $s'$ an $a$-derivative of $s$ iff $s\pijl{a}s'$.

\begin{definition}[$k$-bisimilar \cite{milner1980calculus}]\label{def:k-bisim}
	Let $L = (S, Act,\pijl{})$ be an LTS. For every $k\in \Nat$,
	$k$-bisimilarity written as $\bisim_{k}$ is defined inductively: 
\[	
\arraycolsep=1pt
\begin{array}[t]{rll}
	{\bisim_0} &= \{(s,t) \mid&  s,t\in S\}\text{, and}\\
	{\bisim_{k+1}} &= \{(s,t) \mid& \forall
	s\pijl{a}s'. \exists t\pijl{a}t'\textrm{ such that } s'\bisim_k t',
	\textrm{ and } \\
	& &\forall t\pijl{a} t'. \exists s\pijl{a} s'\textrm{ such that } t'\bisim_k s'\}.
\end{array}
\]
\end{definition}
\noindent Bisimilarity, denoted as ${\bisim}$, is defined as the intersection of
all $k$-bisimilarity relations for all $k\in \Nat$: ${\bisim}=\bigcap_{k\in\Nat}
{\bisim_k}$. As our transition systems are finite, and therefore finitely
branching, $\bisim$ coincides with the more general co-inductive definition of
bisimulation~\cite{park1981concurrency}. The intuition  behind $\bisim_i$ is that
within $i$ (atomic) observations there is no distinguishing behaviour. 
We sketch a rather simple example that showcases this behaviour. 

\begin{figure}
	\centering
	\subcaptionbox{The LTS $\calA_3 = (\{x_0, \dots, x_3\}, \{a\},
	\pijl{})$.\label{fig:depth-minimal}}[0.4\columnwidth]
	{%
	\begin{tikzpicture}[scale=1,
		every node/.style={scale=1},
		node distance = 1.2cm, initial text=]
		\node[] (x3) {$x_3$};
		\node[right= of x3] (x2) {$x_2$};
		\node[right= of x2] (x1) {$x_1$};
		\node[right= of x1] (x0) {$x_0$};

		\path[->]
		(x3) edge node[above] {$a$} (x2)
		(x2) edge node[above] {$a$} (x1)
		(x1) edge node[above] {$a$} (x0);
		%
	\end{tikzpicture}
	}\hfill
	\subcaptionbox{The LTS $\calB_3= (\{x_i,y_i \mid 0\leq i \leq 3\}, \{a\},
		\pijl{})$.\label{fig:neg-depth}}[0.46\columnwidth]
	{%
	\begin{tikzpicture}[scale=0.5,
			every node/.style={scale=1},
			node distance = 1.2cm, initial text=]
			
			\node[] (x3) {$x_3$};
			\node[right= of x3] (x2) {$x_2$};
			\node[right= of x2] (x1) {$x_1$};
			\node[right= of x1] (x0) {$x_0$};
		
			\node[above= of x3] (y3) {$y_3$};
			\node[right= of y3] (y2) {$y_2$};
			\node[right= of y2] (y1) {$y_1$};
			\node[right= of y1] (y0) {$y_0$};
		
			\path[->]
			(x1) edge node [above] {$a$} (x0)
			(x2) edge node [above] {$a$} (x1)
			(x3) edge node [above] {$a$} (x2)
			(y1) edge node [above] {$a$} (y0)
			(y2) edge node [above] {$a$} (y1) 
			(y3) edge node [above] {$a$} (y2)
			(y2) edge node [above right] {$a$} (x1)
			(x3) edge node [above left] {$a$} (y2)
			(x1) edge node [above left] {$a$} (y0) 
			(y0) edge [loop right] node [below] {$a$} (y0);
		\end{tikzpicture}
		}
	\caption{Two example LTSs.\label{fig:examples}}
\end{figure}
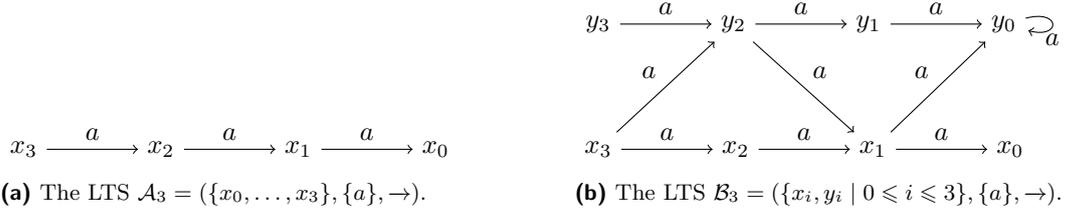

\begin{example}\label{ex:observation-depth}
For every $n\in \Nat$, we define the LTS $\calA_n = (S, \{a\}, \pijl{})$ with a
singleton action set, and the set of states $S=\{x_0, \dots, x_n\}$. The transition
function contains a single path $x_i \pijl{a} x_{i-1}$ for all $1 \leq i\leq n$.

In Figure~\ref{fig:depth-minimal} the LTS $\calA_3$ is shown. A state $x_i$ can
perform $i$ $a$-transitions ending in
a deadlock state. All states in $\calA_3$ are behaviourally inequivalent.
Intuitively, we see that distinguishing the states $x_3$ and $x_2$ takes at
least $3$ observations. 

In general, it holds that for $n\in\Nat$, the states $x_n$ and $x_{n-1}$ of the LTS
$\calA_n$ are $n{-}1$-bisimilar but not $n$-bisimilar, i.e. $x_n \bisim_{n-1}
x_{n-1}$ but $x_n \not\bisim_{n} x_{n{-}1}$. In order to distinguish these
states we require $n$ (atomic) observations. This intuition is formalized in
Theorem~\ref{thm:k-bisim-k-depth}.
\end{example}

\begin{fact}\label{fact:k-distinguishable}%
	We state these well-known facts for an LTS $L= (S, Act, \pijl{})$, and $k\in
	\Nat$:
\begin{enumerate}
\item The relation ${\bisim_{k}}$ is an equivalence relation.
\item If two states are $k$-bisimilar, they are $l$-bisimilar for every $l\leq
  k$.\label{fact:k-distinguishable-1}
\item If ${\bisim_k} = {\bisim_{k+1}}$ then ${\bisim_k} = {\bisim_{k+u}} =
{\bisim}$, for all $u\in\Nat$.\label{fact:k-distinguishable-3}
\end{enumerate}
\end{fact}

For technical reasons we also define $m$-nested
similarity~\cite{groote1992structured} which uses the concept of similarity.
\begin{definition}[Similarity] %
	Given an $L=(S, Act, \pijl{})$, we define \textit{similarity}
	${\simby}\subseteq S\times S$ as the largest relation such that if $s\simby
	t$ then for all transitions $s\pijl{a} s'$ there is a $t\pijl{a}t'$ such
	that $s'\simby t'$.
\end{definition}
\noindent We say a state $s$ is \emph{simulated} by $t$ iff $s \simby t$. 

\begin{definition}[cf. Def. 8.5.2. \cite{groote1992structured}] %
	Let $L=(S,Act,\pijl{})$ be an LTS, and $m\in \Nat$ a number. We inductively
	define $m$-\textit{nested similarity inclusion} as follows: ${\simby^0} = {\simby}$,
	and for every $i\in \Nat$, the relation $\simby^{i{+}1}\subseteq S\times S$ is the
	largest relation such that for all $(s,t) \in {\simby^{i{+}1}}$ it holds that:
	\begin{itemize}
		\item $s\simby^i t$ and $t\simby^i s$, and
		\item if $s\pijl{a}s'$ then there is a $t\pijl{a}t'$ such that
		$s'\simby^{i+1} t'$.
	\end{itemize}
\end{definition}
\noindent We write $\simueq^m$ as the symmetric closure of $m$-nested
similarity inclusion, i.e. ${\simueq^m} = {\simby^m} \cap
\left(\simby^m\right)^{-1}$, which we call $m$-\textit{nested similarity}. Note
that we deviate slightly from the definition in~\cite{groote1992structured},
where $1$-nested simulation equivalence coincides with simulation equivalence.

\begin{example}\label{ex:negdepth} For every $n\in \Nat$, we define the LTS
	$\calB_n = (S, \{a\}, \pijl{})$ with a singleton action set, the set of
	states $S=\{x_0, \dots, x_n, y_0, \dots, y_n\}$, and the transition relation
	containing the transition $y_0 \pijl{a} y_0$ and, for every $i\in [1, n]$,
	the transitions: 
	\begin{itemize} 
		\item $y_i \pijl{a} y_{i-1}$ and $x_i \pijl{a} x_{i-1}$, and 
		\item $y_i \pijl{a} x_{i-1}$ if $i$ is even, or $x_i \pijl{a} y_{i-1}$ if $i$ is odd. 
	\end{itemize}
\end{example}
In Figure~\ref{fig:neg-depth} the LTS $\calB_3$ is shown. We observe that $x_0$
is simulated by $y_0$, since $x_0$ has no outgoing transitions. So it is the
case that $x_0 \simby^0 y_0$, but $y_0 \not\simby^0 x_0$, and hence $x_0
\not\simueq^0 y_0$. In general, for all $n\geq 1$ it holds in the LTS $\calB_n$
that $x_{n} \simueq^{n{-}1} y_n$, but $x_n \not\simueq^{n} y_n$.

\subsection{Hennessy-Milner logic (HML)}
\noindent We use Hennessy-Milner Logic~(HML)~\cite{hennessy1980observing} to
distinguish states. For some finite set of actions $Act$, the syntax of HML is
defined as
\[ \phi ::= \ttrue \mid \langle a \rangle \phi \mid \neg \phi \mid \phi \wedge
\phi\textrm{,}\]
where $a\in Act$. The logic consists of three necessary elements:
\begin{itemize}
\item \emph{Observations $\dia{a}\phi$}, the state witnesses an observation $a$
to a state that satisfies $\phi$. 
\item \emph{Negations $\neg \phi$}, the state does not satisfy $\phi$.
\item \emph{Conjunctions $\phi_1 \wedge \phi_2$}, the state satisfies both
$\phi_1$ and $\phi_2$.
\end{itemize}
The set $\F$ is defined to contain all HML formulas. 
It is common to use the abbreviations $\ffalse=\neg\ttrue$, $[a]\phi=\neg\langle a\rangle\neg\phi$ and
$\phi_1\vee\phi_2=\neg(\neg\phi_1\wedge\neg\phi_2)$. 

Given an LTS
$L = (S,Act,\pijl{})$, we define the semantics of this logic $\sem{-}_L:
\mathcal{F} \rightarrow 2^S$, inductively as follows:
\begin{equation*}
	\label{semantics}
	\begin{split}
	\sem{{\ttrue}}_L &= S\textrm{,}\\
	\sem{\dia{a} \phi}_L &= \{s\in S \mid \exists s'\in S \textrm{ s.t. } s\pijl{a} s'
	\textrm{ and } s'\in \sem{\phi}_L\}\textrm{,}\\
	\sem{\neg \phi}_L &= S \setminus \sem{\phi}_L\textrm{, and}\\
	\sem{\phi_1 \wedge \phi_2}_L &= \sem{\phi_1}_L \cap \sem{\phi_2}_L\textrm{,}
	\end{split}
\end{equation*}
for $a\in Act$ and $\phi, \phi_1, \phi_2\in \mathcal{F}$.  This function yields
for a formula $\phi\in \F$ the subset of $S$ where $\phi$ is true. Often we omit
the reference to the LTS $L$ when it is clear from the context.

We use HML formulas to describe distinguishing behaviour. Let $L=(S,Act,
\pijl{})$ be an LTS, $s\in S$ and $t\in S$ states,  and $\phi\in \F$ a HML
formula. We write $s \sim_{\phi} t$ iff $s\in\sem{\phi} \Leftrightarrow
t\in\sem{\phi}$, and conversely $s\not\sim_{\phi} t$ iff
$s\in\sem{\phi}\Leftrightarrow  t\not\in\sem{\phi}$. Additionally, we write $s
\leq_{\phi} t$ if $s\in\sem{\phi} \Rightarrow t\in\sem{\phi}$. Given a set of
HML formulas $\mathcal{G}$ we write $s \sim_\mathcal{G} t$ iff for every
$\psi\in \mathcal{G}$, it holds that $s\sim_\psi t$. Similarly, we write
$s\leq_\mathcal{G} t$ iff $s\leq_\psi t$ for all $\psi\in \mathcal{G}$.

\begin{definition}
	Given an LTS $L=(S,Act, \pijl{})$ and two states $s,t\in S$, then a formula
	$\phi\in \F$ \textit{distinguishes} $s$ and $t$ iff $s
	\not\sim_\phi t$.
\end{definition}

\subsubsection{Metrics}
\noindent To express the size of a formula we use three different metrics: 
\begin{itemize}
	\item \emph{size} the total number of observations, 
	\item \emph{observation-depth} the largest number of nested observation in
	the formula, and
	\item \emph{negation-depth} the largest number of nested negations in the
	formula.
\end{itemize}
For these metrics we inductively define the functions $\sizef{\_}: \F \to \Nat$ for size,
$\depthf{}: \F \to \Nat$ for observation-depth and $\negdepthf{} : \F \to \Nat$
for negation-depth, as follows:
\[
\arraycolsep=1.2pt
\begin{array}{ll@{\quad}ll@{\quad}ll}
	\size{\ttrue} &= 0,
	& \depth{\ttrue} &= 0,
	& \negdepth{\ttrue} &= 0,\\
	\size{\dia{a}\phi} &= \size{\phi} + 1,
	& \depth{\dia{a}\phi} &= \depth{\phi} + 1,
	& \negdepth{\dia{a}\phi} &= \negdepth{\phi},\\
	\size{\neg \phi} &= \size{\phi},
	& \depth{\neg \phi} &= \depth{\phi},
	& \negdepth{\neg \phi} &= \negdepth{\phi} + 1,\\
	\size{\phi_1 {\wedge} \phi_2} &= \size{\phi_1} {+} \size{\phi_2}.
	&\depth{\phi_1 {\wedge} \phi_2} &= \max(\depth{\!\phi_1\!}, \depth{\!\phi_2\!}). 
	& \negdepth{\!\phi_1 {\wedge} \phi_2\!} &= \max(\negdepth{\!\phi_1\!}, \negdepth{\!\phi_2\!}).
\end{array}
\]



\noindent
Given natural numbers $n,m \in \Nat$ we define the sets $\F_n$ and $\F^m$ as the
fragment of HML formulas with bounded observation- and respectively
negation-depth, i.e. $\F_n = \{ \phi \mid \depth{\phi} \leq n\}$, and $\F^m = \{
\phi \mid \negdepth{\phi} \leq m\}$.

We write $\F^m_n$ for the set $\F^m_n = \F_n \cap \F^m$. Based on these metrics we define multiple notions of \emph{minimal} distinguishing
formulas.

\begin{definition}
	Given an LTS $L= (S, Act, \pijl{})$, let $\phi \in \mathcal{F}$ be an HML
	formula that distinguishes $s\in S$ and $t\in S$. Then in distinguishing $s$
	and $t$, the formula $\phi$ is called:
	\begin{itemize}
		\item to have \emph{minimal observation-depth} iff $\phi$ has the least nested modalities,
		i.e. for all $\phi'\in \F$ if $s\not\sim_{\phi'}t$ then $\depth{\phi}
		\leq \depth{\phi'}$;
    	\item to have \emph{minimal negation-depth} iff $\phi$ has the least nested
    	negations, i.e., for all $\phi'\in \F$ if $s\not\sim_{\phi'}t$ then
    	$\negdepth{\phi} \leq \negdepth{\phi'}$;
		\item to be \emph{minimal} iff $\phi$  has the least number of modalities,
		i.e., for all $\phi'\in\F$ if $s\not\sim_{\phi'}t$ then $\size{\phi}
		\leq \size{\phi'}$;
		\item to have \emph{minimal observation- and negation-depth} iff it is
		minimal in the lexicographical order of observation and negation-depth,
		i.e., iff for all $\phi'\in \F$ if $s\not\sim_{\phi'}t$ then $\depth{\phi}
		\leq \depth{\phi'}$ and if $\depth{\phi} = \depth{\phi'}$ then
		$\negdepth{\phi} \leq \negdepth{\phi'}$;
  		\item \emph{irreducible} \cite[Def. 2.5]{cleaveland1990} iff no $\phi'$
  		obtained by replacing a non-trivial subformula of $\phi$ with the
  		formula $\ttrue$ distinguishes $s$ from $t$.
	\end{itemize}
\end{definition}

\begin{figure}
	\centering

\begin{tikzpicture}[scale=1,
every node/.style={scale=1},
node distance = 1.2cm, initial text=]

\node[] (1) {$s_0$};
\node[right= of 1] (2) {$s_1$};
\node[right= of 2] (3) {$s_2$};

\path[->]
(1) edge node [above] {$a$} (2)
(2) edge node [above] {$a$} (3)
(2) edge [bend right=45] node [above] {$b$} (1);
;

\end{tikzpicture}%
	\caption{The LTS $M=(S, Act, \pijl{})$, $Act=\{a,b\}$ and $S=\{s_0, s_1,
	s_2\}$.\label{fig:cleaveland-minimal}}
\end{figure}
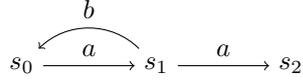

The first three notions correspond directly to the metrics we defined. The
notion of \emph{irreducible} distinguishing formulas corresponds to the
minimality notion used in the work by Cleaveland~\cite{cleaveland1990}. The
different notions are not comparable. This is witnessed by the LTS $M$ pictured
in Figure~\ref{fig:cleaveland-minimal}. The formula $\phi_1 = \dia{a}\dia{a}
\ttrue$ distinguishes $s_0$ and $s_1$ since $s_0 \in \sem{\phi_1}$ and
$s_1\not\in\sem{\phi_1}$. Additionally, $\phi_1$ is irreducible, since any
formula obtained by replacing a subformula by $\ttrue$ is not a distinguishing
formula. However, the formula $\phi_1$ is not \emph{minimal} since the formula
$\phi_2 = \dia{b}\ttrue$ also distinguishes $s_0$ and $s_1$.

\subsubsection{Representation}
\begin{figure}[h]
\centering
\begin{tikzpicture}
\draw (0,0) node (x0) {$\wedge$};
\draw (2,0)  node (shared) {$\langle b\rangle$};
\draw (1,0.5) node (x2) {$\langle a\rangle$};
\draw (3,0) node (x3) {$\langle c\rangle$};
\draw (4,0) node (x4) {$\ttrue$};
\draw[->] (x0) -> (x2);
\draw[->] (x2) -> (shared);
\draw[->] (x0) -> (shared);
\draw[->] (shared) -> (x3);
\draw[->] (x3) -> (x4);
\end{tikzpicture}
\caption{A HML formula represented as a shared term.}
\label{fig:shared_term}
\end{figure}
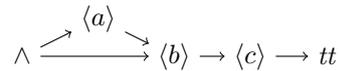

\noindent
A note has to be made on the representation of distinguishing formulas. It is
known that distinguishing formulas can grow very large. In fact there is a
family of LTSs that showcases an exponential lower bound on the size of the
minimal distinguishing formula~\cite{figueira2010size,
wissmann2022quasilinear}. This exponential lower bound is not in
contradiction with the polynomial-time algorithm from
Cleaveland~\cite{cleaveland1990} since \cite{cleaveland1990} uses equations to
represent the subformulas. For example the formula $\langle a\rangle \langle
b\rangle \langle c\rangle \ttrue\wedge \langle b\rangle \langle c\rangle\ttrue$
 \,can be represented using the equations $\phi_1 =\langle a\rangle
\phi_2\wedge\phi_2$ and $\phi_2 =\langle b\rangle \langle c\rangle \ttrue$, or
as the term in Figure \ref{fig:shared_term}. 

The shared representation does not change the observation-depth and the negation-depth. 
The size of a formula is influenced, but it does not affect the NP-hardness result. 
\subsubsection{Correspondences}
\noindent
There are strong correspondences between different fragments of HML on the one hand
and $m$-nested similarity and bisimilarity on the other hand.
We use these to obtain minimal distinguishing formulas. The first theorem states that those
HML formulas that have at most $k$-nested observations exactly capture
$k$-bisimilarity. 

\begin{theorem}{(cf. \cite[Theorem
2.2]{hennessy1980observing})}\label{thm:k-bisim-k-depth}
	Given an LTS $L=(S, Act, \pijl{})$ and two states $s,t\in S$. For every
	$k\in \Nat$, \[s \bisim_k t \iff s \sim_{\F_k} t.\]
\end{theorem}

In this work we are mainly interested in the contraposition of this theorem. For
every $k\in\Nat$, two states $s,t\in S$ are not $k$-bisimilar iff there is a
$\phi \in \F_k$ that distinguishes $s$ and $t$, i.e. $s\not\sim_{\phi} t$. For
this reason for every $k\in \Nat$ we call $s$ and $t$ $k$-\emph{distinguishable}
iff $s\not\bisim_k t$. We call the states $s$ and $t$ \emph{distinguishable} iff
they are $k$-distinguishable for some $k\in \Nat$.

\begin{corollary}\label{cor:k-bisim-k-depth}
Given an LTS $L=(S, Act, \pijl{})$ and two states $s,t\in S$. 
For every $k\in \Nat$,
\[s \not\bisim_k t \iff \text{there is a formula } \phi \in \F_k\text{ such that } s
\not\sim_{\phi} t.\]
\end{corollary}

%


In~\cite{groote1992structured} it is shown that fragments of HML with bounded
negation-depth allow a similar relational classification.
The following
theorem relates the fragment $\negF^m$ to $m$-nested similarity inclusion. 
\begin{theorem}{(cf. \cite[Corollary 8.7.6]{groote1992structured})}\label{thm:negdepth-groote}
	Let $L= (S, Act, \pijl{})$ be an LTS, then for all $m\in \Nat$, and states
	$s,t\in S$: \[s\simby^m t \iff  s \leq_{\negF^m} t.\]
\end{theorem}

The main use for our work is that if two states are not $m$-nested similar, then
there is a distinguishing formula with at most $m$ nested negations. 

\begin{corollary}\label{cor:negdepth-groote}
	Let $L=(S, Act, \pijl{})$ be an LTS, then for all $m\in \Nat$, and states $s,t\in S$:
	\[s \not\simby^m t \iff \text{there is a formula } \phi \in \negF^m\text{ s.t.\ } s\in \sem{\phi}
	\text{ and } t\not\in \sem{\phi}.\]
\end{corollary}

Let us recall the LTS $\calA_3$ from Example~\ref{ex:observation-depth} drawn in Figure~\ref{fig:depth-minimal}. 
In this LTS we see that $x_3 \bisim_2 x_2$,
but $x_3 \not \bisim_3 x_2$. As a result of Corollary~\ref{cor:k-bisim-k-depth} we
know that there is a formula $\phi\in \F_3$ that distinguishes $x_3$ and $x_2$. 
This is witnessed by the formula $\phi =
\dia{a}\dia{a}\dia{a}\ttrue \in \F_3$, which is a distinguishing formula, since
$x_3\in \sem{\phi}$ and $x_2\not\in \sem{\phi}$. We also see that $x_3
\sim_{\F_2} x_2$, hence there is no such formula in $\F_2$.

For the LTS $\calB_3$ from Example~\ref{ex:negdepth}, we aim to distinguish the
states $x_3$ and $y_3$. According Corollary~\ref{cor:negdepth-groote} there is a
distinguishing formula $\phi\in \negF^3$, since $x_3 \not\simby^{3} y_3$. This
is witnessed by the formula $\phi =
\dia{a}\neg\dia{a}\neg\dia{a}\neg\dia{a}\ttrue$. This is a
distinguishing formula as $x_3\in \sem{\phi}$ and $y_3\not\in\sem{\phi}$.
Corollary~\ref{cor:negdepth-groote} also shows that this is the minimal negation-depth
formula distinguishing $x_3$ and $y_3$, as $x_3 \simueq^2 y_3$.

\subsubsection{Traces}
\noindent
Let $Act$ be a finite set of action labels. We denote by $Act^*:= \bigcup_{i\in
\Nat} Act^i$ the set of all finite sequences on the action labels $Act$. We
write $\epsilon$ for the empty sequence. For sequences  $w,u \in Act^*$, we denote
with $|w|$ its length and  $w\concat u$ the concatenation of $w$ and $u$,
which is sometimes also written as $wu$.

\begin{definition}\label{def:trace} Given an LTS $L=(S, Act, \pijl{})$. The set
of \emph{traces} $\traces(s)\subseteq Act^*$ of a state $s\in S$ is the smallest
set satisfying:
  \begin{enumerate}
    \item $\epsilon \in \traces(s)$, and
    \item for an action $a\in Act$, and state $s'\in S$ if a trace $w\in \traces(s')$ and $s\pijl{a}s'$,
    then $aw \in \traces(s)$. \label{def:trace:2}
  \end{enumerate}
\end{definition}
Inductively, we define the formula $\phi_w$ for every word $w\in Act^*$, such
that $\phi_\epsilon = \ttrue$, and $\phi_{aw} = \dia{a} \phi_w$. We call a
formula $\phi\in \F$ a \emph{trace-formula} iff there is a sequence $w\in Act^*$ such
that $\phi = \phi_w$.

\begin{lemma}\label{lem:tracehml}
  Let $L=(S, Act, \pijl{})$ be an LTS, and $w\in Act^*$ a trace. Then for all $s\in S$:
  \[s\in\sem{\phi_w} \iff w\in \traces(s).\]
\end{lemma}
Two states $s\in S$ and $t\in S$ in an LTS $L=(S,Act,\pijl{})$ are said to be
trace-equivalent iff $\traces(s) = \traces(t)$. Bisimilarity is a more
fine-grained equivalence than trace equivalence. Two states $s\in S$ and $t\in
S$ can be trace-equivalent, while not being bisimilar. In this case there is a
formula $\phi\in F$ such that $s\not\sim_\phi t$ and we know that $\phi$ is not
a trace-formula. However, $\phi$ contains traces that are both traces of $s$ and $t$. To
make this more precise we define the traces of a formula by induction for
formulas $\phi, \phi_1,\phi_2\in \F$ as follows:
\[\begin{split}
	\traces(\ttrue) &= \{\epsilon\},\\
	\traces(\dia{a}\phi) &= \{a\} \cup \{a\concat w \mid w \in \traces(\phi) \},\\
	\traces(\neg \phi) &= \traces(\phi),\\
	\traces(\phi_1 \wedge \phi_2) &= \traces(\phi_1) \cup \traces(\phi_2).
\end{split}
\] 
The traces of a formula allow us to state the correspondence between
$k$-distinguishability and the length of shared traces. We formulate this using
the minimal observation depth that, given two distinguishable states, yields the
smallest $i\in\Nat$ such that the states are $i$-distinguishable:
\begin{definition}
Let $L= (S, Act , \pijl{})$ be an LTS. 
We define the \textit{minimal observation depth} $\dist:
S\times S\to \Nat\cup\{\infty\}$ by
$$\dist(s,t) = \left\{
\begin{array}{ll}
i&\textrm{if } s\not\bisim_i t\text{, and } s\bisim_{i-1}t,\\
\infty&\textrm{if }s\bisim t.
\end{array}\right.$$
\end{definition}

The next lemma says that if states have minimal observation depth $i$, then any
distinguishing formula contains a trace of length at least $i$.

\begin{lemma}\label{lem:tracesphi}%
  Let $L=(S, Act,\pijl{})$ be an LTS and $s,t\in S$ two distinguishable states
	such that $\dist(s,t) = i$ for some $i \in \Nat$. For all $\phi\in \F$, if
	$s\not\sim_\phi t$ then there is a trace $w\in \traces(\phi)$ such that $|w|
	\geq i$ and $w\in\traces(s)\cup \traces(t)$.
\end{lemma}
\begin{proof}[Proof sketch.]
	Proven by induction on the shape of $\phi$. The only interesting case is if
	$\phi = \dia{a} \phi'$ for some $a\in Act$ and $\phi'\in \F$. Assume without
	loss of generality that $s\in \sem{\phi}$ and $t\not\in \sem{\phi}$. This
	means that there is a transition $s\pijl{a}s'$ such that $s'\in
	\sem{\phi'}$. Since $\dist(s,t) = i$ there is also a $t\pijl{a}t'$ such that
	$\dist(s', t') = i-1$.
		
	Since $t\not\in\sem{\phi}$ also $t'\not\in \sem{\phi'}$, and thus we can
	apply our induction hypothesis to conclude that there is a $w'\in
	\traces(\phi')$ such that $|w'| \geq i-1$ and $w'\in \traces(s') \cup
	\traces(t')$. From $w'$ we construct $aw'$ and observe that $aw'\in
	\traces(\phi)$, $aw'\in \traces(s) \cup \traces(t)$ and $|aw'| \geq i$,
	which finishes the proof. 
\end{proof}
\section{NP-hardness results}\label{sec:np-hard}
\noindent%
In this section we show that
finding minimal distinguishing formulas is NP-hard. 

We first show that the existence of a short trace is NP-complete similar to a
result of Hunt~\cite[Sec. 2.2]{hunt74} on acyclic NFAs. A corollary of the construction is
that finding the \emph{minimal size} distinguishing formula is NP-hard. 

We define the decision problems \textit{TRACE-DIST} and \textit{MIN-DIST}. Given
an LTS $L=(S, Act, \pijl{})$, two states $s,t\in S$ such that $s\not\bisim_i t$
for $i = |S|$, and a number $l\in \Nat$.
\begin{description}
  \item[\textit{TRACE-DIST}:] There is a trace-formula $\phi\in\F_i$, such that $\phi$
  distinguishes $s$ and $t$.
  \item[\textit{MIN-DIST}:] There is a formula $\phi\in\F_i$, such that $\phi$
  distinguishes $s$ and $t$, and $\size{\phi}\leq l$.
\end{description}
We point out that \textit{TRACE-DIST} is not the same as deciding
trace-equivalence. The problem \textit{TRACE-DIST} decides whether there is a
distinguishing trace of length $i$, and $i$ is smaller than the number of
states, and a minimal distinguishing trace might be super-polynomial in
size~\cite[Sec. 5]{ellul2005regular}. 
\subsection{Reduction}
\noindent%
We prove that \textit{TRACE-DIST} is NP-complete and \textit{MIN-DIST} is
NP-hard by a reduction from the decision problem \textit{CNF-SAT}. This
decision problem decides whether a given propositional formula $\mathcal{C}$ in
conjunctive normal form~(CNF) is satisfiable. For this we define an LTS
$L_\mathcal{C}$, based on the CNF formula $\mathcal{C}$.

\begin{definition}\label{def:reductie-np-hard} Let $\C = C_1
	\wedge \ldots \wedge C_n$ be a CNF formula over the set of proposition letters $\prop =
	\{p_1, \ldots, p_k\}$. We define the LTS $L_\C = (S, Act, \pijl{})$ as follows:
\begin{itemize}
  \item The set of states $S$ is defined as
  \begin{align*}
    S =& \{ \unsat^C_i \mid C \in \{C_1, \ldots, C_n\}, i\in[0,k]\} \cup
	  \{ \sat_i \mid  i\in [0,k]\} \\ 
    & \cup \{\bot_i \mid i \in [0,k]\} \cup
	  \{s,t, \delta\}.
	\end{align*}
  \item The set of actions $Act$ is defined as
    \[
      Act = \{p, \overline{p} \mid p\in \prop\} \cup \{\init, \flag\}.
    \]
	\item The relation ${\pijl{}}$ contains for each $C\in \{C_1, \ldots ,C_n\}$
	and $i\in [1,k]$:
      \[\begin{array}{r@{\hspace{0.1cm}}c@{\hspace{0.1cm}}l}
            \unsat^C_{i-1}&\pijl{p_i}& \left\{
            \begin{array}{ll} 
               \sat_{i}&\textrm{ if }p_i \textrm{ is a literal of } C,\\
               \unsat^C_{i}&\textrm{ otherwise,}
            \end{array}\right.\\
            \unsat^C_{i-1}&\pijl{\overline{p_i}}&
            \left\{\begin{array}{ll} 
            \sat_{i}&\textrm{ if }\neg p_i \textrm{ is a literal of } C,\\
            \unsat^C_{i}&\textrm{ otherwise, }
            \end{array}\right.\\
            \sat_{i-1} &\pijl{x}& \sat_{i}\textrm{ for }x\in \{p_i, \overline{p_i}\}, \textrm{ and}\\
            \bot_{i-1} &\pijl{x}& \bot_{i}\textrm{ for }x\in \{p_i, \overline{p_i}\}.
      \end{array}\]
      Additionally, it contains the auxiliary transitions
      \[\begin{array}{r@{\hspace{0.1cm}}c@{\hspace{0.1cm}}l}
          \unsat^C_{k} &\pijl{\flag}&\delta\textrm{ for } C\in \{C_1, \ldots , C_n\},\\
          \bot_k &\pijl{\flag}& \delta\textrm{,}\\
          t &\pijl{\init}& \unsat^C_0\textrm{ for } C\in \{C_1, \ldots , C_n\},\\
          t &\pijl{\init}& \sat_0\textrm{,} \\
          s &\pijl{\init}& \sat_0\textrm{, and} \\
          s &\pijl{\init}& \bot_0.
        \end{array}\]
  \end{itemize}
\end{definition}

\begin{figure}[t]
  \centering
  \includegraphics[width=0.4\columnwidth]{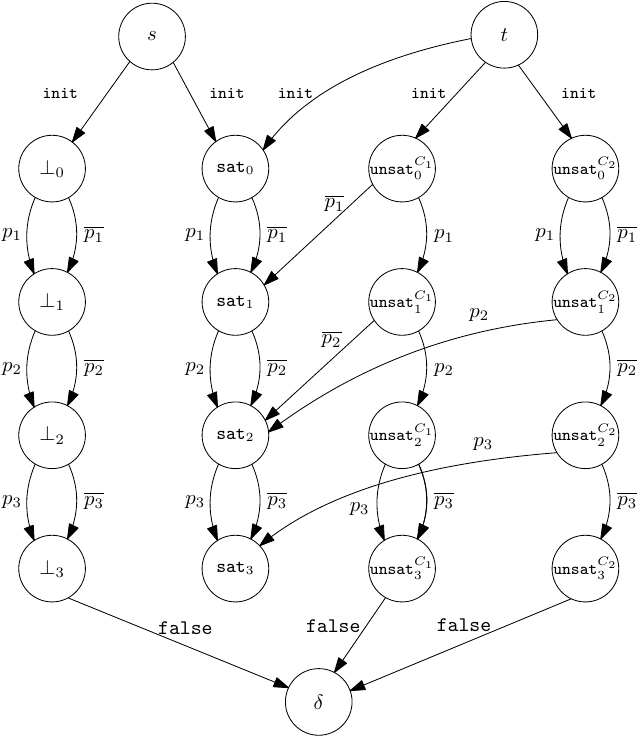}
  \caption{The LTS $L_\mathcal{C}$ for the formula $\mathcal{C} =
  (\neg p_1 \vee \neg p_2) \wedge (p_2 \vee p_3)$.} 
  \label{fig:ltscnf}
\end{figure}

The LTS $L_\mathcal{C}$ for the CNF formula ${\cal C}=C_1\, \wedge\, C_2$ with
clauses $C_1 = \neg p_1 \vee \neg p_2$ and $C_2=p_2 \vee p_3$ is depicted in
Figure \ref{fig:ltscnf}. 

In this construction an interpretation of the propositions $\prop = \{p_1, \ldots
, p_k\}$ is directly related to a word $w = a_1\dots a_k$, where $a_i\in \{p_i,
\overline{p_i}\}$ for every $i\in[1,k]$. The set of truth
assignments encoded as words is defined as:
\[\truths = \{a_1\dots a_k \mid  a_i\in \{p_i,
\overline{p_i}\}\text{ for all } i\in[1,k]\}.\]

Given a truth assignment $\rho: \prop \to \B$, we define $w_\rho$ as $w_\rho =
a_1\dots a_k$, where $a_i = p_i$ if $\rho(p_i)= \textit{true}$ and $a_i =
\overline{p_i}$, otherwise. Conversely, for a word $w=a_1\dots a_k$, a trace
from $\truths$, it represents the truth assignment $\rho_w$ defined for each $i
\in [1,k]$ as:
\[  \rho_w(p_i) = \left\{\begin{array}{ll}
       \textit{true} &\text{if } a_i = p_i,\\
       \textit{false} &\text{if } a_i = \overline{p}_i.
     \end{array}\right.
\]

The idea of the construction of $L_\C$ is that it contains a $\bot$ component, a
$\sat$ component, and an $\unsat^C$ component for every clause $C$. All
components are deterministic and acyclic, and hence describe a finite set of
traces. All the traces of these components start by a truth assignment $w\in
\truths$. By construction, for every truth assignment $w\in\truths$,
$w\concat\flag \in \traces(\bot_0)$. In this way the $\bot$ component represents
falsehood. Conversely, the state $\sat_0$ represents a tautology, since for any
truth assignment $w\in \truths$, $w\concat\flag \not\in\traces(\sat_0)$. For
every clause $C$, and truth assignment $w\in\truths$ the state $\unsat^C_0$
contains $w\concat\flag$ as trace iff $\rho_w$ does not satisfy $C$.

\begin{lemma}\label{lemma:traces} Let $L_\C=(S, Act, \pijl{})$ be the LTS for a CNF formula $\C =
  C_1 \wedge \ldots \wedge C_n$ with propositions $\{p_1, \ldots, p_k\}$, then:
  \[
	\begin{split}
		\traces(\sat_0) &= \{u\in Act^* \mid \exists w\in\truths\text{. } u \text{ is a prefix of } w\} , \\
		\traces(\bot_0) &= \traces(\sat_0) \cup \{w{\cdot}\flag \mid w\in \truths\}, \textrm{ and} \\
		\traces(\unsat^C_0) &= \traces(\sat_0) \cup \{w{\cdot}\flag \mid w\in \truths \text{ and } 
		\rho_w \text{ does not satisfy } C\}.
	\end{split}
	\]
\end{lemma}
\noindent
This lemma is easily verified from the construction of $L_\C$.

\begin{corollary}\label{col:shape} Let $w\in\truths$ be a trace, and $L_\C$ the
  LTS for the CNF formula $\C= C_1\wedge \ldots \wedge C_n$. Then for any clause
  $C\in\{C_1,\ldots,C_n\}$:
    $$w{\concat}\flag \in \traces(\unsat^C_0) \iff C \text{ is not satisfied under
    }\rho_w.$$
\end{corollary}

The following lemma contains the main idea for the reduction of the main theorem
showing \textit{TRACE-DIST} is NP-complete.

\begin{lemma}\label{lem:traceiffsat} Given the LTS $L_\mathcal{C}= (S,
	{\pijl{}}, Act)$ for a CNF formula $\mathcal{C}= C_1 \wedge \ldots \wedge
	C_n$, with propositions $\prop=\{p_1,\ldots, p_k\}$. Then there is a trace
	$w\in Act^{k+1}$ such that $w\in\traces(\bot_0)$, and
	$w\not\in\traces(\unsat_0^C)$ for every $C\in\{C_1,\ldots, C_n\}$ if and only
	if $\mathcal{C}$ is satisfiable.
\end{lemma}
\begin{proof} 
	 We prove this in both directions separately.
	\begin{description}
		\item[$({\Rightarrow})$] As a witness, we obtain a trace $w\in
		\traces(\bottom_0)$ of length at most $k{+}1$ such that $w\not\in
		\traces(\unsat_0^C)$ for all clauses $C\in \{C_1,\ldots, C_n\}$. Since $w
		\in \traces(\bottom_0)$ by Lemma~\ref{lemma:traces} either $w\in
		\traces(\sat_0)$ or $w\in\{v\concat\flag \mid v\in \truths\}$. Since
		$\traces(\sat_0)\subseteq \traces(\unsat^C_0)$, and $w\not\in
		\traces(\unsat^C_0)$, there is a trace $v\in\truths$ such that $w=
		v\concat\flag$. By Corollary~\ref{col:shape} all clauses $C$ are satisfied
		by $\rho_w$. This means $\rho_w$ is a satisfying assignment for $\C$.
     \item[$({\Leftarrow})$] If there is a satisfying assignment $\rho$ for
     $\mathcal{C}$ then we show that $w_\rho\concat \flag$ witnesses the
     implication. First observe that by definition $w_{\rho}\concat \flag \in
     \traces(\bottom_0)$. Let $C\in\{C_1,\ldots ,C_n\}$ be any clause. Since
     $\rho$ is a satisfying assignment, $C$ is satisfied under $\rho$. This
     means by Corollary~\ref{col:shape} that $w_{\rho}\not\in
     \traces(\unsat_0^C)$.\qedhere
	\end{description}
\end{proof}

Now we are ready to prove the main theorem of this section. 

\begin{theorem}\label{thm:trace:np-complete}%
	Deciding TRACE-DIST is NP-complete.
\end{theorem}
\begin{proof}
  First we verify that TRACE-DIST is in NP. Given an LTS $L=(S,Act,\pijl{})$,
  and two states $s,t\in S$. As a witness we get a formula $\phi \in \F_{|S|}$,
  which is a trace-formula. Since $\depth{\phi}\leq |S|$ this is polynomial in
  size. It is well known that given a formula $\phi$ we can check in polynomial
  time whether $s\sim_\phi t$.

	To show TRACE-DIST is NP-hard we reduce CNF-SAT to TRACE-DIST. Let
	$\mathcal{C} = C_1\wedge \ldots \wedge C_n$ be a CNF formula over the
	propositions $\prop = \{p_1, \ldots , p_k\}$. Then for the LTS
	$L_\mathcal{C}$ we show there is a distinguishing trace smaller than $|S|$
	for $s\in S$ and $t\in S$ if and only if $\mathcal{C}$ is satisfiable.

	We begin by observing the sets $\traces(s), \traces(t)$:
	\begin{align*}
		\traces(s) &= \{\epsilon,\init\} \cup{} \{\init \concat w \mid w\in \traces(\bot_0) \cup \traces(\sat_0) \},\\
		\traces(t) &= \{\epsilon,\init\} \cup{} \{\init \concat w \mid w\in \traces(\sat_0) \cup \bigcup_{i\in[1,n]}\traces(\unsat^{C_i}_0)\}.
	\end{align*}

	Since for every $C\in\{C_1,\ldots, C_n\}$, $\traces(\unsat^C_0) \subseteq
	\traces(\bot_0)$ and $\traces(\sat_0) \subseteq \traces(\unsat_0^C)$, we
	know that if there is a distinguishing trace it has to be $\init\concat w\in
	\traces(s)$ for a $w\in \traces(\bot_0)$. By Lemma~\ref{lem:traceiffsat}
	this trace $w$ exists iff $\mathcal{C}$ is satisfiable. Hence, the states
	$s$ and $t$ are in \textit{TRACE-DIST} if and only if $\mathcal{C}$ is in
	\textit{CNF-SAT}. The LTS $L_{\mathcal{C}}$ can be computed in polynomial
	time, as it has $(n+2)(k+1) + 3$ states and $2k(n+2)+2n+4$ transitions. This
	concludes the proof that \textit{TRACE-DIST} is NP-complete.
\end{proof}

In the reduction a distinguishing trace is also a minimal distinguishing
formula. Which means we can generalise our NP-hardness result. 

\begin{corollary}
	Deciding MIN-DIST is NP-hard.\label{thm:dist-np-hard}
\end{corollary}
\begin{proof}

	We prove this by a similar reduction as in the proof of
	Theorem~\ref{thm:trace:np-complete}. The intuition is that, given a CNF
	formula $\C = C_1 \wedge \ldots \wedge C_n$ with propositions $\prop = \{p_1,
	\dots, p_k\}$, in the LTS $L_\C$ a distinguishing formula $\phi\in \F$ such
	that $\size{\phi} = k+2$ necessarily is a trace-formula.

	We reduce CNF-SAT to MIN-DIST. Let $\mathcal{C} = C_1\wedge \ldots \wedge C_n$
	be a CNF formula over the propositions $\prop = \{p_1, \ldots , p_k\}$. Then
	for the LTS $L_\mathcal{C}$ we show there is a distinguishing formula $\phi
	\in \F$ for $s\in S$ and $t\in S$ such that $\size{\phi} \leq k+2$ if and
	only if $\mathcal{C}$ is satisfiable.

	For the direction $\Rightarrow$, assume a formula $\phi \in \F$ exists such
	that $\size{\phi} \leq k+2$ and $s\not\sim_\phi t$. We show that this means
	$\C$ is satisfiable. We observe by the deterministic behaviour
	that $s\bisim_{k+1} t$. Hence, by Theorem~\ref{thm:k-bisim-k-depth} we know
	$\depth{\phi} \geq k+2$. Since we assume $\size{\phi}\leq k+2$ we know that
	$\depth{\phi} = k+2$ and so, there are no non-trivial conjunctions, and we see
	that we can rewrite $\neg\neg \phi \mapsto \phi$. Hence, there is a formula
	$\psi = \triangle_1 \dots \triangle_{k+2} \ttrue$ such that for each $i\in
	[1, k+2]$, $\triangle_i\in \{\dia{a_i}, \neg\dia{a_i}\}$, for some
	$a_1,\ldots,a_{k+2}\in Act$, such that $\sem{\phi} = \sem{\psi}$.
	
	 By Lemma~\ref{lem:tracesphi} there is a trace $w\in \traces(\psi)$, such
	 that $|w|\geq k+2$ and, $w\in \traces(s) \cup \traces(t)$. The only trace
	 of this length of $s$ or $t$ is in the shape $w= \init\concat \hat{p}_1
	 \dots \hat{p}_k\concat \flag$, where $\hat{p_i}\in \{p_i, \overline{p}_i\}$
	 for each $i\in [1,k]$. This means that $a_1 = \init$, $a_{j+1} = \hat{p}_j$
	 for each $j\in [1,k]$ and $a_{k+2} = \flag$. We are going to show that the
	 associated truth value $\rho = \rho_{\hat{p}_1\dots\hat{p}_k}$ satisfies
	 $\C$ by reductio ad absurdum.  
	 	 
	 If $\rho$ does not satisfy $\C$ then there is a clause $C$ such that $C$ is
	 not satisfied by $\rho$. We claim for this clause $\unsat_0^C
	 \sim_{\Delta_2\ldots\Delta_{k+1}\ttrue} \bot_0$, and since both $s$ and $t$
	 have a $\init$-transition to $\sat_0$ this means $\psi$ does not distinguish
	 any of the derivatives. Hence $s\sim_{\psi} t$ which is a contradiction. 

	For the other direction if $\C$ is satisfiable then by
	Lemma~\ref{lem:traceiffsat} there is a $w\in Act^{k+1}$ such that $w\in
	\traces(\bot_0)$ and $w\not\in \traces(\unsat^C_0)$ for all clauses $C\in
	\{C_1,\ldots, C_n\}$. Using $w$ we construct the distinguishing trace $w' =
	\init\concat w$. Since $w\in \traces(\bot_0)$, $w\not\in
	\traces(\unsat^C_0)$  and by construction also $w\not\in\traces(\sat_0)$, it
	is the case that $w' \in \traces(s)$ and $w'\not\in \traces(t)$. This means
	the formula $\phi_{w'}$ is a distinguishing formula and $\size{\phi_{w'}} =
	k+2$, which finishes the second part of the proof. 
\end{proof}

The problem MIN-DIST is not a member of NP since a polynomially sized witness
might not exist. However, there is always a `shared' distinguishing formula of
polynomial size. Since we can compute in polynomial time if a shared formula is
a distinguishing formula, the decision problem MIN-DIST formulated in terms of
total `shared' modalities is NP-complete.
\section{Efficient algorithms}\label{sec:dist}
\noindent%
In this section we explain that despite the NP-hardness results from the
previous section it is still possible to efficiently generate distinguishing
formulas with minimal observation- and negation-depth. First, we introduce the
method $\phi(s,t)$ listed in Algorithm~\ref{lst:phi_naive} that generates a
minimal observation-depth distinguishing formula for the states $s$ and $t$. We
extend $\phi(s,t)$ to the function $\psi_i(s,t)$ listed in
Algorithm~\ref{lst:phi}. This method computes a distinguishing formula with
observation-depth of at most $i$ and minimal negation-depth. Additionally, this
procedure also prevents unnecessary conjuncts to be added. Finally, we indicate
how to compute the equivalences $\bisim_1, \dots, \bisim_k$, and the minimal
observation- and negation-depth.  
\subsection{The algorithm}
\noindent%
For every $i\in \Nat$, we define a function $\splpairs_i :
S\times S \to 2^{Act \times S}$ that gives all distinguishing observations. More
precisely, given two $i$-distinguishable states $s\in S$ and $t\in S$,
$\splpairs_i(s,t)$ returns all pairs $(a,s')$, where $a\in Act, s'\in S$, such
that $s\pijl{a}s'$ and $s'$ is $(i{-}1)$-distinguishable from all targets
$t\pijl{a}t'$. The definition of $\splpairs_i(s,t)$ is: 
\[ 
  \splpairs_i(s,t) = \{(a,s')\mid s\pijl{a}s' \text{ and }\forall t\pijl{a} t'.\ \dist(s',t') \leq i-1 \}.
\]

Using the function $\splpairs_i(s,t)$, we can compute a minimal
observation-depth formula using the procedure listed as
Algorithm~\ref{lst:phi_naive}. The procedure selects an action state pair
$(a,s')\in \splpairs_i(s,t)$ and recursively distinguishes $s'$ from all
$a$-derivatives of $t$. If $\splpairs_i(s,t)$ is empty the negated $\phi_i(t,s)$
is calculated and in this case $\splpairs_i(t,s)$ is necessarily not empty.

\begin{algorithm}[t]
  \SetKwInOut{Input}{input}
  \SetKwInOut{Output}{output}
  \Input{Two states $s,t\in S$ such that $s\not\bisim_i t $}
  \Output{A formula $\phi\in \F$ s.t. $s\in \sem{\phi}$ and $t\not\in \sem{\phi}$}

  \SetKwProg{Fn}{Function}{ is}{end}
  \Fn{$\phi(s,t)$}{
    $i := \dist(s,t)$\;
    \uIf{$\splpairs_i(s,t) = \emptyset$} {
      \KwRet{$\neg\, \phi(t,s)$}
    }
    Select $(a,s')\in \splpairs_i(s,t)$\;
    $T := \{t' \mid t\pijl{a} t'\}$\;
    \KwRet{$\dia{a} \left(\, \bigwedge_{t'\in T} \phi(s',t') \,\right)$}\;
  }
  \caption{Minimal-depth distinguishing formula\label{lst:phi_naive}}
\end{algorithm}

\begin{lemma}\label{lemma:exists-split}%
  Given an LTS $L=(S, Act, \pijl{})$ and two states $s,t \in S$. If
  $s\not\bisim_i t$ then:
  $\splpairs_i(s,t) \neq \emptyset \text{ or } \splpairs_i(t,s) \neq
  \emptyset.$
\end{lemma}
\begin{proof}
  As $s\not\bisim_i t$ there either is an $s\pijl{a}s'$ such
  that $s'\not\bisim_{i-1} t'$ for all $t\pijl{a}t'$, or vice-versa there is a
  $t\pijl{a}t'$ such that $t'\not\bisim_{i-1}s'$ for all $s\pijl{a}s'$. In the
  first case $(a,s')\in \splpairs_i(s,t)$, in the second case $(a,t')\in
  \splpairs_i(t,s)$.
\end{proof}

\subsection{Minimal negation-depth}
\noindent In order to minimize the number of negations within the minimal
observation-depth formula we combine the notions of $k$-bisimilar and $m$-nested
similarity inclusion.

\begin{definition}\label{def:simbyaux} Let $L=(S,Act \pijl{})$ be an LTS, and
	$k,m\in \Nat$. We define $m$-nested $k$-similarity inclusion, denoted
	$\simbyaux_k^m$, inductively by for all $s,t\in S$, $s\simbyaux^m_0 t$ and if $s\simbyaux^m_k t$
	then
	\begin{enumerate}
		\item if $s\pijl{a}s'$ there is a $t\pijl{a}t'$ such that
		$s'\simbyaux^m_{k-1} t'$, and
		\item if $m>0$ and $t\pijl{a}t'$, then there is a $s\pijl{a}s'$ such that
		$t'\simbyaux^{m-1}_{k-1} s'$.\label{def:simbyaux:2}
	\end{enumerate}
\end{definition}

Similarly to the original Hennessy-Milner correspondences, we observe the
correspondence between the fragment $\negF_k^m$ and the relation ${\sba{k}{m}}$.
\begin{theorem}\label{thm:sba}
  Let $L=(S,Act, \pijl{})$ be an LTS. For any $k,m\in \Nat$ and states $s,t\in S$:
  \[s\leq_{\negF_k^m} t \iff s\sba{k}{m} t.\]
\end{theorem}
\noindent%
Related to the distance measure $\dist$, we define the directed
minimal negation-depth measure for the relation ${\sba{k}{m}}$, for states that are
not $m$-nested $k$-similar for some $k, m\in \Nat$. 
\begin{definition}\label{def:minimal-negation-depth}
  Let $L= (S, Act , \pijl{})$ be an LTS and $i\in \Nat$ be a number. 
  We define the \textit{directed minimal negation-depth} $\dirdistsingle_i:
  S\times S\to \Nat\cup\{\infty\}$ by
  \[\dirdistsingle_i(s,t) = \left\{
  \begin{array}{ll}
  j&\textrm{if } s\not\sba{i}{j} t\text{, and } s\sba{i}{j{-}1} t,\\
  \infty&\textrm{if }s\bisim_i t.
  \end{array}\right.\]
\end{definition}
For every $i,j\in \Nat$ we define a function $\hat{\splpairs}^j_i: S\times S \to 2^{Act \times S}$ that is
similar to the function $\splpairs_i$. It adds an extra limitation on the number
of negations needed to distinguish the pairs from all observations from $t$.   
\[ 
  \hat{\splpairs}^j_i(s,t) = \{(a,s')\mid (a,s')\in \splpairs_i(s,t) \text{ and }\forall t\pijl{a} t'.\ \dirdistsingle_{i-1}(s',t') \leq j \}.
\]%
The next lemma guarantees that a suitable distinguishing observation exists. 

\begin{lemma}\label{lemma:exists-split:neg}%
  Given an LTS $L=(S, Act, \pijl{})$ and two states $s,t \in S$. Then for all
  $i,j\in \Nat$, if $s\not\sba{i}{j} t$ then $
    \hat{\splpairs}^j_i(s,t) \neq \emptyset \text{ or } \hat{\splpairs}^{j-1}_i(t,s) \neq
  \emptyset.$
\end{lemma}
\begin{algorithm}[t]
  \SetKwInOut{Input}{input}
  \SetKwInOut{Output}{output}
  \Input{Two states $s,t\in S$ such that $s\not\bisim_{i} t $ for some $i \in \Nat$}
  \Output{A formula $\phi \in \negF_{i}$ such that $s\in \sem{\phi}$ and $t\not\in \sem{\phi}$}
  \SetKwProg{Fn}{Function}{ is}{end}
  \Fn{$\phi_i(s,t)$}{
    $j := \dirdistsingle_i(s,t)$\;
    $\X := \hat{\splpairs}^j_i(s,t)$\;
    \uIf{$\X = \emptyset$} {
      \KwRet{$\neg\, \phi_i(t,s)$}
    }
    Select $(a,s')\in \X$\;
    $T := \{t' \mid t\pijl{a} t'\}$\;
    \While{$T\neq \emptyset$}{
      Select $t_{\mathit{max}} \in T$ such that $\dirdistsingle_{i-1}(s', t_{\mathit{max}}) \geq \dirdistsingle_{i-1}(s', t')$ for all $t'\in T$\;
      $\phi_{t_{\mathit{max}}} := \phi_{i-1}(s', t_{\mathit{max}})$\;
      $\Phi := \Phi \cup \{\phi_{t_{\mathit{max}}}\}$\;
      $T := T \cap \sem{\phi_{t_{\mathit{max}}}}$\;
    }
    \KwRet{$\dia{a} \left(\, \bigwedge_{\phi\in \Phi} \phi\right)$}
  }
  \caption{Generate a distinguishing formula with minimal observation- and negation-depth.\label{lst:phi}}
\end{algorithm}

In Algorithm~\ref{lst:phi} we give the method $\psi_i(s,t)$ that given an LTS
$L=(S, Act, \pijl{})$ and $i$-distinguishable states $s,t\in S$ generates a
formula such that $s\in\sem{\psi_i(s,t)}$ and $t\not\in \sem{\psi_i(s,t)}$ with
observation depth at most $i$ and minimal negation-depth.

The algorithm attempts to find an action label $a\in Act$ and an $a$-derivative
$s\pijl{a} s'$, such that all $a$-derivatives $t'$, such that $t\pijl{a}t'$ are
distinguishable with a formula with at most $i{-}1$ nested observations and $j$
nested negations. These pairs $(a,s')$ are given by the function
$\hat{\splpairs}^j_i(s,t)$. In Line~6 one of these witnesses is chosen. If there
is more than one suitable derivate, one is chosen at random. 

The next theorem states that Algorithm \ref{lst:phi} yields a valid
distinguishing formula.
\begin{theorem}\label{thm:psi-splits} %
  Let $L=(S,Act, \pijl{})$ be an LTS, and $s, t\in S$ be states. If $s$ and $t$ are
  $k$-distinguishable for some $k\in\Nat$ then $s\in \sem{\psi_k(s,t)}$ and
  $t\not\in\sem{\psi_k(s,t)}$.
\end{theorem}

The next theorem states that if $\dist(s,t) = k$, then $\psi_k(s,t)$ yields a
formula that has minimal observation-depth, and there is no formula $\phi$ with a
smaller number of nested negations such that $s\not\leq_\phi t$. 
\begin{theorem}\label{lem:psi-minimal-depths}%
  Let $L=(S,Act, \pijl{})$ be an LTS, and $s, t\in S$ be states, such that
  $s\not\bisim t$ and $\dist(s,t) = k$. Then for all $\phi\in\F$, if $s
  \not\leq_\phi t$ then $\depth{\psi_k(s,t)} \leq \depth{\phi}$ and if
  $\negdepth{\phi} < \negdepth{\psi_k(s,t)}$ then $\depth{\phi} >
  \depth{\psi_k(s,t)}$.
\end{theorem}

\noindent%
\subsection{Partition refinement}
\noindent
In order to execute Algorithm~\ref{lst:phi}, we need to compute the functions
$\dist$ and $\dirdistsingle$. In this section we propose a simple partition
refinement algorithm that does exactly this by first computing the relations
$\bisim_0, \bisim_1, \dots, \bisim_k$ iteratively. The pseudocode is listed in
Algorithm~\ref{lst:seq-partition-refinement}. In contrast to the more efficient
partition refinement algorithms~\cite{hopcroft1971DFAmin, paige1987three,
valmari2009bisim}, we guarantee that \emph{older} blocks are used first as
splitter. This method is inspired by~\cite{smetsers2016minimal} where pairwise
minimal distinguishing words are computed.

Most algorithms deciding bisimilarity are so-called partition refinement
algorithms~\cite{kanellakis1983,paige1987three}. Our algorithms are also based
on partition refinement. A \textit{partition} $\partit$ of a set $S$ is a
disjoint cover of $S$, i.e. a set of non-empty subsets of $S$ and every element
of $S$ is in exactly one subset. The elements $B\in \partit$ are called
\textit{blocks}. A partition $\partit$ induces the equivalence relation
$\sim_{\partit}:S\times S$ in which the blocks are the equivalence classes, i.e.
$\sim_{\partit} = \{(s,t) \mid \exists B\in \pi \text{ and } s,t\in B\}$. In the
algorithm we filter a set of states $U$ on a distinguishing observation with
respect to a set of given states $V$, and an action $a\in Act$, i.e.: $\spl_a(U,
V) = \{s \in U \mid \exists s'\in V. s\pijl{a}s'\}$.

\setlength{\textfloatsep}{1.5em}
\begin{algorithm}[t]
  \SetAlgoNoEnd
	\SetKwProg{Fn}{Function}{ is}{end}
	\SetKwFunction{FRefine}{$\mathtt{Refine}$}
	\Fn{\FRefine{$\pi$}}{
		$\pi' := \pi$\;
		\ForEach{$a\in Act, B'\in \pi$\label{line:loop1}}
			{
				\ForEach{$B\in\pi'$\label{line:loop2}}{
					$C := \spl_a(B,B')$\;\label{line:split:1}
					\If{$C \neq B$ and $C\neq \emptyset$}{
						$\pi' := (\pi' \setminus \{B\}) \cup \{C, B\setminus C\}$\; \label{line:split:2}
					}
				}
			}
			\KwRet{$\pi'$}\;
	}
  $i:= 0$; $\pi_0:=\{S\}$\;\label{line:initpi0}
  \While{$\pi_{i} \neq \FRefine(\pi_{i})$}{
    $\pi_{i+1} := $ \FRefine{$\pi_i$}\;\label{line:refinepi}
    $i := i+1$\;
  }
	\caption{Iterative partition refinement.\label{lst:seq-partition-refinement}}
\end{algorithm}

The next theorem states that the procedure listed as
Algorithm~\ref{lst:seq-partition-refinement} produces a sequence of partitions,
in which the $i$-th partition induces $i$-bisimilarity.
\begin{theorem}\label{thm:correctness-partition-refinement}%
  Given an LTS $L=(S,Act,\pijl{})$ and partitions $\pi_0, \dots, \pi_k$ produced
  by Algorithm~\ref{lst:seq-partition-refinement}. Then ${\sim_{\pi_i}} =
  {\bisim_i}$, for all $0\leq
  i\leq k$.
\end{theorem}

\noindent 
It is possible to compute the function~$\dirdistsingle_i(s,t)$ in polynomial
time from the computed $k$-bisimilarity relations calculated in
Algorithm~\ref{lst:seq-partition-refinement}. It is important to use dynamic
programming such that $\dirdistsingle_i(s,t)$ for every $i$, $s$ and $t$ is only
calculated once. 

\subsection{Evaluation}
\noindent%
The computation of Algorithm~\ref{lst:phi} needs to account for redundancies to
guarantee a polynomial time algorithm. We use dynamic programming to achieve
this. For any pair of states $s,t\in S$ if the function $\psi_i(s,t)$ is invoked, it stores the generated shared formula. Whenever the
function is called again, the previously generated formula is used, with only
constant extra computing and memory usage. Hence, given an LTS $L=(S,Act,
\pijl{})$ the number of recursive calls is limited to the combination of states and level $k\leq |S|$,
i.e. $\bigO(|S|^3)$ calls.

\begin{corollary}
  Given an LTS $L=(S,Act, \pijl{})$ and a pair of distinguishable states $s,t\in
  S$, then the following is computable in polynomial time:
  \begin{itemize}
    \item A minimal observation-depth distinguishing formula,
    \item A minimal observation- and negation-depth distinguishing formula. 
  \end{itemize}
\end{corollary}

A naive implementation of the algorithms requires quadratic memory. This could
be a bottleneck for large state spaces. Representing the equivalences $\bisim_k$
as a splitting tree~\cite{lee1994testing} is more memory efficient. In addition,
an optimization is to generate only distinguishing formulas between equivalence
classes of the generated equivalences, instead of individual states.


\begin{table}[b]
    \centering
    \begin{adjustbox}{max width=0.95\textwidth}
        \begin{tabular}{l|r|r|r|r|r|r|r|r|r|r|r|r|r}
        & \multicolumn{6}{c|}{Max}
        & \multicolumn{6}{c|}{Average}\\
        \hline
        \multirow{ 2}{*}{Benchmark} & 
        \multicolumn{2}{c|}{$\depth{\phi}$} &
        \multicolumn{2}{c|}{$\size{\phi}$} &
        \multicolumn{2}{c|}{$\negdepth{\phi}$} &
        \multicolumn{2}{c|}{$\depth{\phi}$} &
        \multicolumn{2}{c|}{$\size{\phi}$} &
        \multicolumn{2}{c|}{$\negdepth{\phi}$} \\\cline{2-13}
    & Our & Cleav. & Our & Cleav. & Our & Cleav. & Our & Cleav. &  Our & Cleav. & Our & Cleav.
    \\
    \hline
    ieee-1394-1 & 64 & 891 & 69 & 1355 & 0 & 886 & 64,0 & 247,2 & 69,0 & 373,7 & 0,0  & 243,2  \\ 
    ieee-1394-2 & 37 & 224 & 42 & 320 & 1 & 219 & 37,0 & 92,0 & 42,0 & 120,0 & 1,0  & 88,2  \\ 
    ieee-1394-3 & 102 & 698 & 102 & 1092 & 2 & 696 & 102,0 & 299,1 & 102,0 & 465,4 & 2,0  & 295,7  \\ 
    ieee-1394-4 & 76 & 363 & 83 & 506 & 2 & 360 & 76,0 & 196,6 & 80,9 & 276,5 & 2,0  & 194,5  \\ 
    ieee-1394-5 & 18 & 155 & 18 & 214 & 2 & 146 & 18,0 & 36,0 & 18,0 & 44,8 & 2,0 & 30,4 \\ 
\end{tabular}
\end{adjustbox}
\caption{\label{tab:results}Results from prototype implementation Algorithm 1.}
\end{table}%


We implemented a prototype of the method introduced here. We also implemented
the method proposed by Cleaveland~\cite{cleaveland1990} in which we decided
bisimilarity by a partition refinement algorithm in which the splitter selected
is the latest created block, since heuristically this has the best runtime
\cite{baclet2006around, berkholz2017tight}. For Cleaveland's method the strategy
for splitter selection matters for the size of the formulas generated. However,
regardless of strategy chosen, the formulas that our method generates are always
more concise in all metrics.
   
We post-processed the formulas to ensure both implementations resulted in
formulas that are irreducible. For the benchmark we used the model
from~\cite{luttik97p1394} containing $188.568$ states and $340.607$ transitions.
We compared this model to $5$ modified versions where we omitted one randomly
chosen transition.
In Table~\ref{tab:results} the results of running the algorithms $10$ times are
shown. Under `Max', the worse-case of the different runs for each metric is
listed for our method (`Our'), next to the result of the implementation of
Cleaveland (`Cleav.'). Under `Average' the average of the $10$ runs is shown.

We see that our new method consistently outputs a minimal observation- and
negation-depth formula, and the generated formulas only rarely deviates in size.
It outperforms the method of Cleaveland in all cases. In some cases the depth is
improved a factor $10$.


\section{Conclusions \& Future work}
In this work we studied the problem of computing minimal distinguishing
formulas. We introduced three metrics: size, observation-depth, and
negation-depth. Using a reduction directly from CNF-SAT we showed that finding a
minimal sized distinguishing formula is NP-hard. However, for observation- and
negation-depth, we introduce polynomial time algorithms that compute minimal
formulas. A prototype demonstrates the potential improvement over the method
introduced by Cleaveland~\cite{cleaveland1990}. A more rigorous version is
implemented in the mCRL2 toolset~\cite{mcrl2}. 

For future work it would be interesting to extend our algorithms for
equivalences beyond strong bisimilarity. For instance, a more generic
coalgebraic treatment, extending~\cite{wissmann2022quasilinear}, or computing
smaller witnesses for equivalences with abstractions like branching and weak
bisimilarity, improving upon the work of Korver~\cite{korver1991computing}. 

\bibliographystyle{plainurl}
\bibliography{bibliography}
\appendix
\newpage
\section{Proofs of Section~\ref{sec:prelims}\label{appendix:prelims}}
\begin{proof}[Proof of Theorem~\ref{thm:k-bisim-k-depth}]
$(\Longrightarrow)$ Given $\phi \in \mathcal{F}$ we show by structural induction on
 the shape of $\phi$ that if $s\bisim_i t$ for some $i$ and $\phi\in \F_i$ then $s\in\sem{\phi} \iff t\in \sem{\phi}$,
 and hence, $\phi$ does not distinguish $s$ and $t$. For the base case if $\phi = \ttrue$,
 this means that $\sem{\phi} = S$ and trivially $s \sim_\phi t$.  For the
 inductive case assume $s,t\in S$ such that $s\bisim_i t$ for some $i$, and $\phi \in \F_i$.
 We distinguish the following cases, in which $\phi_1,\phi_2\in \mathcal{F}$ are 
 smaller formulas for which the induction hypothesis hold.
\begin{itemize}
\item If $\phi = \phi_1 \wedge \phi_2$ then if $s\in\sem{\phi}$, by definition
	$s\in \sem{\phi_1}$ and $s\in\sem{\phi_2}$. By our induction hypothesis it
	must also be the case that $t\in \sem{\phi_1}$ and $t\in \sem{\phi_2}$, and
	so, $t\in\sem{\phi}$. Since $s$ and $t$ are chosen arbitrarily this holds also
	in the other direction and we conclude  
	$s\sim_\phi t$.
\item 
    If $\phi = \neg \phi_1$ then $s\in \sem{\phi}$ iff $s\not\in \sem{\phi_1}$. 
	By the induction hypothesis $s \in \sem{\phi_1}\iff t\in \sem{\phi_1}$.  Hence, $s\in\sem{\phi} \iff t\in\sem{\phi}$.
\item 
	If $\phi = \dia{a} \phi_1$ for some $a\in Act$ then if $s\in \sem{\phi}$ there
	must be an $s'\in S$ such that $s\pijl{a}s'$ and $s'\in \sem{\phi_1}$. Since
	$s\bisim_i t$, by Definition~\ref{def:k-bisim} there is a $t'\in S$ such that
	$t\pijl{a} t'$ and $s'\bisim_{i-1} t'$. In addition, since  $\phi \in \F_i$,
	by definition $\phi_1 \in \F_{i-1}$. Hence our induction hypothesis applies,
	and we derive that $s' \in \sem{\phi_1} \iff t'\in \sem{\phi_1}$, and since
	$s'\in\sem{\phi_1}$, also  $t'\in \sem{\phi_1}$. This witnesses that
	$t\in\sem{\phi}$. Since $s,t$ are chosen arbitrarily we can conclude that
	$s\in\sem{\phi} \iff t\in\sem{\phi}$.
\end{itemize}
 $(\Longleftarrow)$ We prove the other direction using induction on $n$. The
induction hypothesis is that for some $i\in \Nat$ and $s,t\in S$, if it holds that $s\sim_{\F_i} t$ then
$s\bisim_i t$.

For the base case, when $i=0$, the property is trivially true since all states $s,t\in S$ are related, i.e.\ $s\bisim_0 t$.

Now we prove that the property holds for $i{+}1$. Assume, to arrive at a
contradiction, that the property does not hold for $i{+}1$. Then there are two
states $s,t\in S$ such that $s\sim_{\F_{i{+}1}} t$ and $s\not\bisim_{i+1}t$. By
definition~\ref{def:k-bisim} and we chose $s,t$ arbitrarily, we can assume
without loss of generality that there is a state $s'\in S$ such that
$s\pijl{a}s'$ and for all $t'\in S$ such that $t\pijl{a}t'$ we have that
$s'\not\bisim_i t'$.

Now consider the set of target states $T =\{t'\in S \mid t\pijl{a} t' \}$. For
any $t'\in T$ since $s'\not\bisim_i t'$ it follows by our induction hypothesis
that there is a formula $\phi' \in \F_i$ such that $s'\not\sim_{\phi'} t'$. We
define $\phi_{t'} := \phi'$ if $s'\in\sem{\phi_{t'}}$, and $\phi_{t'} := \neg
\phi'$ if $s'\not\in\sem{\phi}$. This means that for all $t'\in T$ we have that
$s'\in\sem{\phi_{t'}}$ and $t'\not\in\sem{\phi_{t'}}$. These formulas allow us
to construct the formula $\phi_{\lightning} = \dia{a} (\bigwedge_{t' \in T}
\phi_{t'})$. We observe that $\phi_{\lightning}\in\F_{i+1}$, while also
$s\in\sem{\phi_{\lightning}}$ and $t\not\in\sem{\phi_{\lightning}}$. This causes
a contradiction since we assumed $s\sim_{\F_{i{+}1}} t$. Hence, it must be true
that the property holds.
\end{proof}

\begin{proof}[Proof of Lemma~\ref{lem:tracehml}]
Proven by induction on $w$. If $w= \epsilon$, then $\phi_w = \ttrue$ and
trivially $\epsilon \in \traces(s)$ and $s\in \sem{\ttrue}$ for all $s\in S$.

The induction hypothesis is that for any $w\in Act^*$ and $s\in S$ it holds that
$s\in\sem{\phi_w} \iff w\in \traces(s)$. We show that for all $a\in Act$ the
property holds for $aw$.

We show both implications simultaneously. Assume $s\in \sem{\phi_{aw}}$ (resp.
$aw\in\traces(s)$) then by definition there is a transition $s\pijl{a}s'$ such
that $s'\in \sem{\phi_w}$ (resp. $w\in\traces(s')$). By our induction hypothesis
we know that $w\in \traces(s') \iff s'\in \sem{\phi_w}$. Therefore,
$w\in\traces(s')$ (resp. $s'\in\sem{\phi_w}$) and by definition
$aw\in\traces(s)$ (resp. $s\in\sem{\phi_{aw}}$). This proves both implications
and concludes the proof.
\end{proof}

\begin{proof}[Proof of Lemma~\ref{lem:tracesphi}]
  We prove this by induction on the shape of $\phi$. For the base case if $\phi
  = \ttrue$ then $\phi$ never distinguishes states.
  
  Let $\phi\in \F$ be a formula such that $s\not\sim_\phi t$, and as induction
  hypothesis we have that the  property holds for all smaller formulas $\phi_1,
  \phi_2 \in \F$. We distinguish on the shape of $\phi$. If $\phi = \neg \phi_1$
  or $\phi = \phi_1 \wedge \phi_2$ the property trivially holds since the
  induction hypothesis applies for $\phi_1$ and $\phi_2$, which leaves us with the
  only interesting case $\langle a\rangle\phi_1$. 
  
  If $\phi = \dia{a} \phi_1$ for some $a\in Act$, then either there is a $s\pijl{a} s'$
  such that $s'\in \sem{\phi_1}$, or a transition $t\pijl{a} t'$
  such that $t'\in \sem{\phi_1}$. Both cases go completely analogue, 
  so without loss of generality assume there is a transition $s\pijl{a}s'$ such
  that $s'\in \sem{\phi_1}$.
  
  Since $\dist(s,t) = i$, we know that $s\bisim_{i-1} t$ and hence there is a transition
  there is a $t\pijl{a}t'$ such that $\dist(s',t')
  > i-2$, but because $s\not\bisim_{i}t$ it also holds that $\dist(s', t') \leq i-1$.
  Hence we can conclude that $\dist(s',t') = i-1$.
  
  We apply our induction hypothesis on the pair $s', t'$ with the
  distinguishing formula $\phi_1$. This means there is a trace $w'\in
  \traces(\phi_1)$ such that $|w'| \geq i-1$, and  $w'\in \traces(s') \cup
  \traces(t')$. Using $w'$ we construct the sequence $w = aw' \in \traces(\phi)$.
  We see that $w\in \traces(s)\cup \traces(t)$ and $|w|\geq i$ finishing the
  proof.
  \end{proof}

\section{Proofs of Section~\ref{sec:np-hard}\label{appendix:np-hard}}
\begin{proof}[Proof of Lemma~\ref{lemma:traces}]
  We begin by proving the following property with reverse induction. For every
  $i\leq k$:
\begin{align*}
   \traces(\sat_i) &=\{a_{i+1}\dots a_k \mid a_j \in
    \{p_{j},\overline{p}_{j}\} \text{ for } i < j \leq k \},\\
    \traces(\bot_i) &=\{a_{i+1}\dots a_k{\cdot}\flag \mid a_j \in
    \{p_{j},\overline{p}_{j}\} \text{ for } i < j \leq k \}.
  \end{align*}
    As base case we have that $\traces(\bot_k) = \flag$ and $\traces(\sat_k) = \emptyset$.
    Given a number $i\in \Nat$ such that $0\leq i \leq k$, we assume as
    induction hypothesis that the property holds for $i+1$.

    By definition of $L_\C$, the states $\sat_i$ and $\bot_i$ have only two
    outgoing transitions $\sat_i\pijl{a} \sat_{i+1}$, and $\bot_i\pijl{a}
    \bot_{i+1}$ for $a\in\{p_i, \overline{p}_{i+1}\}$. By our induction
    hypothesis $\traces(\sat_i) = \{a_{i+1}\dots a_k \mid a_j \in
    \{p_{j},\overline{p}_{j}\} \text{ for all } i < j \leq k \}$ and
    $\traces(\bot_i) = \{a_{i+1}\dots a_k{\cdot}\flag \mid a_j \in
    \{p_{j},\overline{p}_{j}\} \text{ for all } i < j \leq k \}$. This proves
    the property. 

    Now consider some clause $C\in\{C_1,\dots, C_n\}$. Then we state the
    property that for every $i\in [0, k]$ it holds that $w\in
    \traces(\unsat^C_k)$ iff $w$ is a prefix of $a_{i+1}\dots a_{k}$, where
    $a_j\in\{p_j,\overline{p}_j\}$ for every $j\in[i+1,k]$. Additionally,
    $a_{i+1}\dots a_k{\cdot}\flag \in \traces(\unsat^C_k)$ iff there is no $a_j$
    for which the truth value will satisfy $C$. Then as base case
    $\traces(\unsat^C_k) = \{ \flag \}$, which is true by definition. Let $i \in
    [0,k-1]$ be a number such that the property holds for $\unsat^C_{i+1}$. We
    show that the property holds for $\unsat^C_{i}$. Since $i<k$ we know by
    definition that $\unsat^C_{i}$ has one outgoing $p_{i+1}$ transition
    $\unsat^C_{i}\pijl{p_{i+1}} t_1$ and one outgoing transition
    $\unsat^C_{i}\pijl{\overline{p}_{i+1}} t_2$ for $t_1,t_2 \in \{\sat_i,
    \unsat_i^C\}$, and thus $\traces(\unsat^C_i) = \{p_{i+1}\concat w\mid w\in
    \traces(t_1)\} = \{\overline{p}_{i+1}\concat w\mid w\in \traces(t_2)\}$. We
    case distinguish on the values of $t_1$ and $t_2$.
    \begin{itemize}
    	\item If the target $t_1$ (resp.\ $t_2$) is $\sat_i$, then $p_{i+1}$ ($\neg
    	p_{i+1}$) is a literal of $C$ and thus this truth value satisfies $C$, and
    	$\traces(\sat_i)$ confirms that $\flag$ does not appear in the traces.
    	\item If the target $t_1$ (resp.\ $t_2$) is $\unsat^C_i$, then $p_{i+1}$
    	($\neg p_{i+1}$) is \emph{not} a literal of $C$. By our induction
    	hypothesis $w{\cdot}\flag\in\traces(\unsat^C_i)$ iff there is a
    	proposition $p_j$ for $j\in [i{+}2, k]$ which satisfies $C$. Hence the
    	property holds.
   	\end{itemize}
    This completes the induction and confirms that:
	\[
		\pushQED{\qed} 
		\traces(\unsat^C_0) = \traces(\sat_0) \cup \{w{\cdot}\flag \mid w\in \truths \text{ and } \rho_w \text{ does not satisfy } C\}.
		\qedhere
		\]%
\end{proof}

\section{Proofs of Section~\ref{sec:dist}\label{appendix:dist}}
We prove Theorem~\ref{thm:sba} in two directions using two lemmas.
\begin{lemma}\label{lemma:nested:onlyif} Let $L=(S,Act,\pijl{})$ be an LTS,
	$s,t\in S$ two states, and $m,k\in\Nat$ two numbers. Then the following holds
	for all formulas $\phi\in \F$:
	\[\phi\in \negF^m_k \text{ and } s\simbyaux_k^m t \implies s\leq_{\phi} t.\]
\end{lemma}
\begin{proof}
  The proof uses structural induction on $\phi$. For $\phi=\ttrue$ the property
  holds trivially, since $s\leq_{\ttrue} t$ is always true. For the induction
  case assume $\phi\in \negF^m_k$ and $s\simbyaux_k^m t$. We distinguish on the
  possible shapes of $\phi$ where $\phi_1,\phi_2\in \F$ are smaller formulas for
  which the induction hypothesis holds.
  \begin{itemize}
    \item If $\phi = \dia{a} \phi_1$ then $\phi_1 \in \negF^m_{k-1}$. If $s\in \sem{\phi}$, then there is an $s\pijl{a}s'$ such that $s'\in \sem{\phi_1}$. Since $s\simbyaux_k^m t$, there is by definition a $t\pijl{a}t'$ such that $s'\simbyaux_{k-1}^m t'$. By our induction hypothesis since $\phi_1\in\negF^m_{k-1}$, we know that $s'\leq_{\phi_1} t'$ and since $s'\in\sem{\phi_1}$ also $t'\in\sem{\phi_1}$. This means the transition $t\pijl{a}t'$ witnesses $t\in\sem{\phi}$ and therefore $s\leq_{\phi}t$.
    \item Consider $\phi = \phi_1 \wedge \phi_2$. If $s\in\sem{\phi}$ then  
    $s\in\sem{\phi_1}$ and $s\in\sem{\phi_2}$ by definition. 
    By the induction hypothesis $t\in\sem{\phi_1}$ and $t\in\sem{\phi_2}$, and 
    thus $t\in\sem{\phi}$ implying $s\leq_\phi t$.
    \item In the last case if $\phi = \neg \psi$, then $\psi\in \negF_k^{m-1}$. Since $s\simbyaux_k^m t$ also $t\sba{k}{m-1} s$ by Proposition~\ref{prop:sba-reverse}, and this means  $t\leq_\psi s$ by applying the induction hypothesis. If $s\in\sem{\phi}$ then $s\not\in \sem{\psi}$, and because $t\leq_\psi s$, also necessarily $t\not\in\sem{\psi}$. This means $t\in\sem{\phi}$ and hence $s\leq_\phi t$.
  \end{itemize}
  This finalizes the induction, and proves that if $\phi\in \negF_k^m$ and $s\sba{k}{m} t$ then $s\leq_\phi t$.
\end{proof}
In the next lemma we show the implication in the other direction.
\begin{lemma}\label{lemma:nested:if}
	Let $L=(S,Act,\pijl{})$ be an LTS, and $k,m\in \Nat$ be two natural numbers. The following holds for all states $s,t\in S$:
	\[s\leq_{\negF_k^m} t \implies s\sba{k}{m} t.\]
\end{lemma}
\begin{proof}
	We prove this by induction on $k$. In the base case where $k=0$ the property 
	holds trivially because for all $m\in \Nat$, ${\sba{0}{m}} = S\times S$.

	The induction hypothesis is that for some $k\in \Nat$ it holds for all $m\in \Nat$ that:
 		 \[ s\leq_{\negF_{k}^{m}} t \implies s\sba{k}{m} t.\]
  We prove that the property also holds for $k+1$.

  Assume to arrive at a contradiction that $s\leq_{\negF_{k+1}^m} t$ but $s\not\sba{k+1}{m} t$. Since $s\not\sba{k+1}{m} t$ at least one of the two cases of Definition~\ref{def:simbyaux} does not apply. We distinguish these two cases.
  \begin{itemize}
    \item In the first case, there is an $s\pijl{a} s'$ such that for all $t'\in
    S$ if $t\pijl{a}t'$ then $s'\not\sba{k}{m} t'$. Let $T=\{t' \mid t\pijl{a}t'
    \}$ be the set of $a$-derivatives of $t$. By induction
    $s'\not\leq_{\negF_{k}^m} t'$ for every $t'\in T$, since $s'\not\sba{k-1}{m}
    t'$. Hence, there is a formula $\phi_{t'}\in\negF_{k}^m$ such that
    $s'\not\leq_{\phi_{t'}} t'$. We construct $\phi = \dia{a}(\bigwedge_{t'\in T}
    \phi_{t'})$ and since for every $t'\in T$, $s'\in\sem{\phi_{t'}}$ and
    $t'\not\in \sem{\phi_{t'}}$ we have that $s\in \sem{\phi}$, and
    $t\not\in\sem{\phi}$. Since $\phi\in \negF_{k+1}^{m}$, this is in
    contradiction with our assumption that $s\leq_{\negF_{k+1}^m} t$.
\item 
    For the second case, $m> 0$ and there is a
    $t\pijl{a}t'$ such that for all $s'\in S$ if $s\pijl{a}s'$ then
    $t'\not\sba{k}{m-1}s'$. From here the argument is similar to the first
    case. 
    Define $T = \{s' \mid
    s\pijl{a}s'\}$. By applying the induction hypothesis and the fact that
    $t'\not\sba{k}{m-1} s'$,  we deduce that $t'\not\leq_{\negF_{k}^{m-1}} s'$.
    This means for each $s'\in T$ there is a formula $\phi_{s'}\in
    \negF_{k}^{m-1}$ such that $t'\not\leq_{\phi_{s'}} s'$. We construct the
    formula $\phi = \neg\dia{a}(\bigwedge_{s'\in T} \phi_{s'})$, and see that $\phi
    \in \negF_{k+1}^m$. Since for all $s'\in T$, $t'\in \sem{\phi_{s'}}$ and
    $s'\not\in \sem{\phi_{s'}}$, this also means $s\in\sem{\phi}$ and $t\not\in
    \sem{\phi}$. This would witness $s \not\leq_\phi t$, which is a
    contradiction.

  \end{itemize}
  Both cases lead to a contradiction. The only assumption we made
  that could lead to this contradiction is $s\not\sba{k+1}{m}t$. This means 
  that $s\sba{k+1}{m} t$ finishing the proof.
\end{proof}

With these two lemmas we are ready to prove Theorem~\ref{thm:sba}.

\begin{proof}[Proof of Theorem~\ref{thm:sba}]
 By Lemma~\ref{lemma:nested:if} we know $s\leq_{\negF_k^m} t \implies
 s\sba{k}{m} t$. For the reverse implication if $s\sba{k}{m} t$ and $\phi\in\negF_k^m$
 then by Lemma~\ref{lemma:nested:onlyif} $s\leq_{\phi}t$. Hence $s \sba{k}{m} t
 \implies s\leq_{\negF_k^m} t$, which completes the proof.
\end{proof}

\begin{proof}[Proof of Lemma~\ref{lemma:exists-split:neg}]
  Since $s\not\sba{i}{j} t$, by Definition~\ref{def:simbyaux} there either is an $s\pijl{a}s'$ such
  that $s'\not\sba{i-1}{j} t'$ for all $t\pijl{a}t'$, or the second item of the definition is not true
  and there is a $t\pijl{a}t'$ such that $t'\not\sba{i-1}{j-1} s'$ for all $s\pijl{a}s'$.
  
  In the first case $(a,s')\in \splpairs_i(s,t)$ and $\dirdistsingle_{i-1}(s',t') \leq j$ for all $t\pijl{a}t'$. 
  Therefore, $(a,s')\in \hat{\splpairs}_i^j(s,t)$. In the second case $(a,t')\in
  \splpairs_i(t,s)$, and $\dirdistsingle_{i-1}(t',s') \leq j{-}1$ for all $s\pijl{a}s'$,
  and hence $(a, t') \in \hat{\splpairs}_i^{j-1}(t,s)$. 

  This concludes that in both cases either $\hat{\splpairs}_i^j(s,t) \neq \emptyset$ or
  $\hat{\splpairs}_i^{j-1}(t,s) \neq \emptyset$ which was to be proven.
\end{proof}

\begin{proof}[Proof of Theorem~\ref{thm:psi-splits}]
We prove this for all $k\in\Nat$ by induction. If $k=0$ this is trivial since
there are no states $s$ and $t$ that are $0$-distinguishable. The induction
hypothesis is that for some $k\in \Nat$, for all states $s,t\in S$, if $s$ and
$t$ are $k$-distinguishable then $s\in\sem{\psi_k(s,t)}$ and
$t\not\in\sem{\psi_k(s,t)}$. With this hypothesis we show that the property also
holds for $k{+}1$.

Consider states $s,t\in S$ such that $s$ and $t$ are $(k{+}1)$-distinguishable. 
We distinguish on whether the condition $\X =\emptyset$ in line~4 is true, in
the process of calculating $\psi_{k+1}(s,t)$.
\begin{itemize}
  \item If $\X \neq \emptyset$, then a pair $(a,s')\in \X$ is selected such that
  by definition $s\pijl{a}s'$ and for all $t\pijl{a}t'$, $s'\not\bisim_{k}t'$.
  Hence, we know $s'$ and every $t'\in T$ is $k$-distinguishable. By our
  induction hypothesis this means for every $t'\in T$ that $s'\in
  \sem{\psi_k(s', t')}$ and $t'\not\in\sem{\psi_k(s', t')}$. This means that
  after executing the while-loop Line~8-13, that for all $t'\in T$, $t'\not\in
  \sem{\bigwedge_{\phi\in\Phi}}$. Thus, we can conclude that $s\in
  \sem{\psi_{k+1}(s,t)}$ and $t\not\in\sem{\psi_{k+1}(s,t)}$.
  \item If $\X = \emptyset$, then $\psi_{k+1}(s,t) = \neg \psi_{k+1}(t,s)$. 
  We can see that when executing $\psi_{k+1}(t,s)$ the set calculated in line~4 is not
  empty, i.e.\ $\X\neq \emptyset$.  \\
  By definition of $\dirdistsingle_{k+1}$ we know that $s\not\sba{k+1}{j}t$
  where $j=\dirdistsingle_{k+1}(s,t)$. Since in executing $\psi_{k+1}(s,t)$ we
  got $\X=\emptyset$ the first condition in Lemma~\ref{lemma:exists-split:neg}
  is not true and hence the second holds and $\hat{\splpairs}^{j-1}_{k+1}(t,s) \neq
  \emptyset$ and there is a $t\pijl{a}t'$ such that for all $s\pijl{a}s'$ we
  have $\dirdistsingle_{k}(t',s') \leq j{-}1$. By construction
  $\dirdistsingle_i(t,s) \geq \dirdistsingle_i(s,t){-}1 \geq j{-}1$ and thus
  when executing $\psi_{k+1}(t,s)$ we know that $\X \neq \emptyset$. This means
  that in executing $\psi_{k+1}(t,s)$ the first case of this case distinction
  applies, and $t\in\sem{\psi_{k+1}(t,s)}$ and $s\not\in\sem{\psi_{k+1}(t,s)}$.
  Since $\psi_{k+1}(s,t) = \neg\, \psi_{k+1}(t,s)$, a direct result is that
  $s\in\sem{\psi_{k+1}(s,t)}$ and $t\not\in\sem{\psi_{k+1}(s,t)}$.
\end{itemize}
This completes the induction step that if $s$ and $t$ are
$(k{+}1)$-distinguishable then $s\in\sem{\psi_{k+1}(s,t)}$ and
$t\not\in\sem{\psi_{k+1}(s,t)}$, completing the proof.
\end{proof}

\begin{proof}[Proof of Theorem~\ref{lem:psi-minimal-depths}]
First we prove this by induction for all $k \in \Nat$, if $\dist(s,t) = k$ then
$\depth{\psi_k(s,t)} \leq k$ and $\negdepth{\psi_k(s,t)}\leq \dirdistsingle_k(s,t)$. For the
base case $k=0$ this is trivial since there are no states such that
$\dist(s,t) = 0$.

For the induction case, let the property hold for some $i\in \Nat$. Let $s,t\in
S$ be two states such that $\dist(s,t) = i{+}1$ and
$\dirdistsingle_{i{+}1}(s,t) = j$ for some $j\in \Nat$. We show that
$\depth{\psi_{i+1}(s,t)} \leq i{+}1$ and $\negdepth{\psi_{i+1}(s,t)}\leq j$. We do a case
distinction on the condition $\X = \emptyset$ in line~4. 

In the first case assume $\X \neq \emptyset$ in line~4. Then, also in line~6, a
pair $(a,s') \in \X$ is selected. For this pair it holds that $(a,s') \in
\delta_{i{+}1}(s,t)$ and for all $t'\in T = \{t'' \mid t\pijl{a}t''\}$ that
$\dirdistsingle_i(s',t') \leq j \leq \dirdistsingle_{i{+}1}(s,t)$. Additionally,
since $(a,s')\in \hat{\delta}^j_{i{+}1}(s,t)$, we know $\dist(s',t') \leq i$ for
all $t'\in T$. By our induction hypothesis we conclude that
$\depth{\psi_i(s',t')} \leq i$ for all $t'\in T$ and since after the while-loop
from Line~8-13, the set $\Phi$ only contains the formulas $\psi_i(s',t')$ for
some targets $t'$ in line~14, we know $\depth{\psi_{i+1}(s,t)} \leq i{+}1$.
Additionally, since $(a,s')\in \X$ we know that $\dirdistsingle_{i{+}1}(s',t')
\leq j$ for all $t'\in T$. Hence, by induction $\negdepth{\psi_i(s',t')}\leq j$
and also $\negdepth{\psi_{i+1}(s,t)}\leq j$. 

For the other case assume $\X = \emptyset$. So, $\psi_{i+1}(s,t) = \neg\,
\psi_{i+1}(t,s)$. In this case $\depth{\psi_{i+1}(s,t)} =
\depth{\psi_{i+1}(t,s)}$ and $\negdepth{\psi_{i+1}(s,t)} =
\negdepth{\psi_{i}(t,s)} + 1$. Since $\dirdistsingle_{i{+}1}(s,t)=j$ and $\X =
\emptyset$, we know by definition of $\dirdistsingle_{i{+}1}(s,t)$ that
$\dirdistsingle_{i{+}1}(t, s) + 1 \leq \dirdistsingle{(s,t)}$. This means
$\dirdistsingle_{i{+}1}(t,s) \leq j{-}1$. In executing $\psi_{i+1}(t,s)$ the first
case of this case distinction applies, and we conclude by our induction
hypothesis that $\depth{\psi_{i+1}(t,s)}\leq i{+}1$ and $\negdepth{\psi_{i+1}(t,s)} \leq
j{-}1$. Since $\psi_{i+1}(s,t) = \neg\, \psi_{i+1}(t,s)$ we know $\depth{\psi_{i+1}(s,t)} \leq
i{+}1$ and $\negdepth{\psi_{i+1}(s,t)} \leq j$.

Next we show that for a pair of states $s,t\in S$ such that $\dist(s,t) = k$
that for any $\phi\in \F$ if $s \not\leq_\phi t$ then $\depth{\psi_{k}(s,t)}
\leq \depth{\phi}$, and if $\negdepth{\phi} < \negdepth{\psi_k(s,t)}$ then
$\depth{\phi} > \depth{\psi_k(s,t)}$. 

Assume $\phi\in \F$ such that $s \not\leq_\phi t$. Since $\dist(s,t) = k$ we
know by Definition $s\bisim_{k-1}t$. Hence, by Theorem~\ref{thm:k-bisim-k-depth}
we know that $s\sim_{\F_{k-1}} t$ which means $\phi\not\in \F_{k-1}$, and thus
$\depth{\phi} \geq k$. Since we have shown that $\depth{\psi_k(s,t)} \leq k$,
this means $\depth{\psi_k(s,t)} \leq \depth{\phi}$. Next we show that if
$\negdepth{\phi} < \negdepth{\psi_k(s,t)}$ then $\depth{\phi} >
\depth{\psi_k(s,t)}$. Assume $\negdepth{\phi} < \negdepth{\psi_k(s,t)}$ and
$\dirdistsingle_k(s,t) = j$, then by Definition~\ref{def:minimal-negation-depth}
we know that $s\sba{k}{j-1}t$. By Theorem~\ref{thm:sba} this means that
$s\leq_{\negF_{k}^{j-1}} t$, and thus $\phi\not\in\F_{k}^{j-1}$. Since we
assumed $\negdepth{\phi} < j$, it has to be the case that $\depth{\phi} > k$
which proves $\depth{\phi} > \depth{\psi_k(s,t)}$. 
\end{proof}


\begin{algorithm}[ht]
  \SetKwInOut{Input}{input} \SetKwInOut{Output}{output} \Input{An LTS $L= (S,
  Act, \pijl{})$, and partitions $\pi_1, \dots , \pi_k$ such that ${\sim_{\pi_i}}
  = {\bisim_i}$ for all $i\in[1,k]$.} 

	\SetKwProg{Fn}{Function}{ is}{end}
  \Fn{$\dist(s,t)$}{
    \ForEach{$i\in [1,k]$}{
      \If {$s\not\sim_{\pi_i}t$} {
        \KwRet{$i$}\;
      }
      \KwRet{$\infty$}\;
    }
  }
    \Fn{$\dirdistsingle_i(s,t)$}{ 
      \If{$s\sim_{\pi_i} t$}{
        \KwRet{$\infty$}
      }
     $\X := \{\max(\{0\} \cup \{\dirdistsingle_{i{-}1}(s', t') \mid t\pijl{a}t'\}) \mid (a,s') \in \splpairs_i(s,t)\}$\;
     $\overline{\X} := \{\max(\{1\} \cup \{\dirdistsingle_{i{-}1}(t', s') + 1 \mid s\pijl{a}s'\})\mid (a,t') \in \splpairs_i(t,s)\}$\;
     \KwRet{$\min(\X \cup \overline{\X})$}\;
    }
  \caption{The functions $\dist(s,t)$ and $\protect\dirdistsingle_i(s,t)$ for all $s,t\in S$
  such that $s\not\bisim t$.\label{lst:dist-implementation}}
\end{algorithm}

In order to show the correctness of
Algorithm~\ref{lst:seq-partition-refinement} we first state a lemma expressing that $k$-bisimilar states are
not split by equivalence classes of $(k{-}1)$-bisimilarity.
\begin{lemma}\label{lem:never-split}%
  Let $L=(S,Act,\pijl{})$ be an LTS. Given two states $s\in S$ and $t\in S$ such
  that $s\bisim_i t$ for some $i\in \Nat$. Then for all $a\in Act$, blocks $B\in
  \classes{\bisim_{i-1}}$, and sets $U\subseteq S$ such that $s,t\in U$, it holds that
  $$s\in\spl_a(U, B) \iff t\in\spl_a(U,B).$$
\end{lemma}
\begin{proof}
  Assume $s\in \spl_a(U,B)$, for a set of states $U\subseteq S$ such that $s,t
  \in U$, an action $a\in Act$ and a block $B\in\classes{\bisim_{i-1}}$. We show
  that $t\in\spl_a(U,B)$.

  By definition of $\spl_a$, and since $s\in\spl_a(U,B)$, we know that there is
  a state $s'\in B$ such that $s\pijl{a}s'$. Since $s\bisim_i t$, we know by
  Definition~\ref{def:k-bisim} that there is a $t'$ such that $t\pijl{a}t'$ and
  $s'\bisim_{i-1} t'$. Since $B\in \classes{\bisim_{i-1}}$ and $s'\in B$, also
  $t'\in B$ and therefore also $t\in \spl_a(U,B)$. By symmetry, since $s$ and
  $t$ are chosen arbitrarily, $t\in\spl_a(U,B)$ also implies $s\in\spl_a(U,B)$.
  This concludes the proof.
\end{proof}

Now we are ready to prove Theorem~\ref{thm:correctness-partition-refinement}.
\begin{proof}[Proof of Theorem~\ref{thm:correctness-partition-refinement}.]
	We prove this by induction on $i$. For the base case if $i=0$ this is trivial as
$\pi_0=\{S\}$ is assigned in line~\ref{line:initpi0}, and ${\sim_{\pi_0}} =
{\bisim_0}$. For the induction step we assume for a number $i\in \Nat$ such that
$i<k$ that ${\sim_{\pi_i}} = {\bisim_i}$ and show that ${\sim_{\pi_{i+1}}} =
{\bisim_{i+1}}$. This is the same as saying that for all $s,t\in S$ it holds
that $s \sim_{\pi_{i+1}} t \iff s\bisim_{i+1} t$. We show this in both
directions separately for $\pi_{i+1} = \mathtt{Refine}(\pi_i)$.
\begin{description}
  \item[$s \sim_{\pi_{i+1}} t \Leftarrow s\bisim_{i+1} t$:] Assume
  $s\bisim_{i+1} t$, then by
  Fact~\ref{fact:k-distinguishable} item \ref{fact:k-distinguishable-1} also
  $s\bisim_i t$. By the induction hypothesis we know that $s \sim_{\pi_i} t$. In
  other words there is a $B\in\pi_i$ such that $s,t\in B$. If at some stage in
  $\mathtt{Refine}(\pi_i)$ a set $U\subseteq S$ is split with respect to an
  action $a\in Act$ and a block $B'\in \pi_i$ in
  lines~(\ref{line:split:1}-\ref{line:split:2}), then $U$ is divided into two
  sets $C=\spl_a(U, B')$ and $U\setminus C$. By Lemma~\ref{lem:never-split} if
  $s,t\in U$ then either $s,t \in C$ or $s,t \in U\setminus C$. Thus $s$ and $t$
  are not split. Since $s\sim_{\pi_i}t$ and there is no stage in which $s$ and
  $t$ are split into different blocks there is also a block $B\in \pi_{i+1}$
  such that $s,t\in B$ and this means $s\sim_{\pi_{i+1}} t$.
  \item[$s\sim_{\pi_{i+1}} t \Rightarrow s\bisim_{i+1} t$:] 
  Assume $s\sim_{\pi_{i+1}}t$, and $s\pijl{a} s'$ for some $a\in Act$ and $s'\in
  S$. We show there is a $t'\in S$ such that $t\pijl{a} t'$ and $s'\bisim_i t'$.
  There is a block $B_{s'}\in\pi_i$ such that $s' \in B_{s'}$. Since $\pi_{i+1}
  = \mathtt{Refine}(\pi_i)$, the loop from line~\ref{line:loop1} is at some
  point executed with the variables $B_{s'}$ and $a$. Since we assumed $s
  \sim_{\pi_{i+1}} t$ at all iterations of $\mathtt{Refine}$ there is a block
  $B\in \pi'$ such that $s,t \in B$. For this block the set $C:= \spl_a(B,
  B_{s'})$ is calculated in line~\ref{line:split:1}. Since we assumed
  $s\pijl{a}s'$ and $s'\in B_{s'}$ we know $s\in C$. Because $t$ is not split
  from $s$ we also know that $t\in C$ and by the definition of $\spl_a$ there is
  a $t'\in B_{s'}$ such that $t\pijl{a}t'$. By the induction hypothesis we know
  that ${\sim_{\pi_i}} = {\bisim_i}$, and, hence, that $t' \bisim_i s'$. Since the states
  $s,t$ and the transition $s\pijl{a}s'$ are chosen arbitrarily, by symmetry for
  any transition $t\pijl{a}t'$ there is also a transition $s\pijl{a}s''$ such
  that $t'\bisim_{i} s''$. This proves $s\bisim_{i+1} t$ and concludes the
  proof.\vspace{-2.4ex}
\end{description}
\end{proof}

In Algorithm~\ref{lst:dist-implementation} it is shown how to compute the
function~$\dirdistsingle_i(s,t)$ from the computed $k$-bisimilarity relations
calculated in Algorithm~\ref{lst:phi}. In the next Lemma we prove that the
function $\dirdistsingle_i(s,t)$ described in
Algorithm~\ref{lst:dist-implementation} is equal to
Definition~\ref{def:minimal-negation-depth}.

First let us state the following proposition.
\begin{proposition}\label{prop:sba-reverse} Let $L=(S,Act,\pijl{})$ be an LTS.
	For all $k,m\in \Nat$, if $m\geq 1$ and $s\sba{k}{m} t$ then $t\sba{k}{m-1} s$.
\end{proposition}

\begin{lemma}\label{lem:dirdist-sba} Let $L=(S, Act, \pijl{})$ be an LTS. Then
  for all $i\in [0,|S|]$ and states $s,t\in S$ it holds for $\dirdistsingle_i$ from
  Algorithm~\ref{lst:dist-implementation} that either: 
  \begin{itemize}
    \item $s\bisim_i t$ and $\dirdistsingle_i(s,t) = \infty$, or
    \item $s\not\bisim_i t$, and $\dirdistsingle_i(s,t) = j$ for some $j\in \Nat$, such
    that $s\not\sba{i}{j} t$ and $s\sba{i}{j{-}1}t$.
  \end{itemize}
\end{lemma}
\begin{proof}
  We prove this property by induction on $i$. If $i=0$ then $s\bisim_0 t$ for
  all states $s,t\in S$. Since also $s \sim_{\pi_0}t$ for all $s,t\in S$, the
  function yields $\dirdistsingle_0 = \infty$ in line~11. 
  
  For the inductive case assume the property holds for $i$, and let $s,t\in S$
  be states. We show that the Lemma holds for $i{+}1$ and the states $s$ and
  $t$. First assume $s\bisim_{i{+}1} t$. In this case by
  Theorem~\ref{thm:correctness-partition-refinement} also $s\sim_{\pi_{i+1}} t$
  and $\dirdistsingle_{i{+}1}(s,t) = \infty$ in line~11. 
    
  Now assume $s\not\bisim_{i{+}1} t$. We first show that in this case
  $\dirdistsingle_{i{+}1}(s,t) = j$ such that $s\not\sba{i}{j}t$. By
  Lemma~\ref{lemma:exists-split} we know that $\X\cup \overline{\X} \neq
  \emptyset$, and $\dirdistsingle_{i{+}1}(s,t) = j = \min(\X\cup
  \overline{\X})$. We distinguish whether $j\in\X$ or $j\in\overline{\X}$. 
  \begin{itemize}
    \item In the first case if $j \in \X$,  then there is a pair $(a,s')\in
  \delta_{i+1}(s,t)$, such that $\dirdistsingle_{i{+}1}(s', t') = l \leq j$ for all
  $t\pijl{a}t'$. Hence, by applying our induction
  hypothesis, this means $s' \not\sba{i}{l} t'$, and, by contraposition of
  Proposition~\ref{prop:sba-reverse}, in particular $s' \not\sba{i}{j} t'$.
  This means that the first condition of Definition~\ref{def:simbyaux} is not
  satisfied and hence $s\not\sba{i+1}{j} t$. 
     \item In the other case if $j\in \overline{\X}$ pair $(a,t')\in
    \delta_{i+1}(t,s)$ such that for all $s\pijl{a}s'$, we have
    $\dirdistsingle_{i}(t', s')= l \leq j-1$. Hence, by our induction hypothesis $t'
    \not\sba{i}{l} s'$. In particular this means $t' \not\sba{i}{j-1} s'$ which
    contradicts Definition~\ref{def:simbyaux}.\ref{def:simbyaux:2} so
    $s\not\sba{i+1}{j} t$.
  \end{itemize}
  To complete the proof we show that $s\sba{i+1}{j{-}1}t$, by showing both
  properties of Definition~\ref{def:simbyaux}. 
	\begin{enumerate}
		\item For a transition $s\pijl{a}s'$, since $\dirdistsingle_{i+1}(s,t) = j$, there
		is a transition $t\pijl{a}t_{\mathit{max}}$ such that either $s'\bisim_i
		t_{\mathit{max}}$, which witnesses $(a,s') \not\in \delta_{i+1}(s,t)$, or
		$\dirdistsingle_{i}(s',t_{\mathit{max}}) = l \geq j$ by definition of
		$\dirdistsingle_{i+1}$. In the first case it follows that $s'\sba{i}{j-1}
		t_{\mathit{max}}$, by definition. In the second case it holds by the induction
		hypothesis that $s'\sba{i}{l-1} t_{\mathit{max}}$ for $l \geq j$.
		This also means $s'\sba{i}{j-1} t'$.
		\item If $j > 0$ then for a transition $t\pijl{a}t'$ a similar story holds.
		Since $\dirdistsingle_{i+1}(s,t) = j$ then for every transition
		$t\pijl{a}t'$ there is a $s\pijl{a}s_{\mathit{max}}$ such that
		$s_{\mathit{max}}\bisim_i t'$ witnessing $(a,t') \not\in
		\delta_{i{+}1}(t,s)$ or $\dirdistsingle_{i}(t',s_{\mathit{max}}) + 1\geq j$.
		In the first case $t'\sba{i}{j-1}s_{\mathit{max}}$ by definition, and in the
		other case by our induction hypothesis $t'\sba{i}{j-2}s_{\mathit{max}}$.
		Hence, the second property holds.
	\end{enumerate}
	We have shown that both requirements from Definition~\ref{def:simbyaux} hold,
	hence we conclude $s\sba{i+1}{j-1}t$.
\end{proof}

\end{document}